\documentclass[a4paper,10pt,eqno]{article}
\usepackage{theorem}
\usepackage{latexsym,amssymb,amsfonts,amsmath}
\usepackage{graphicx}
\newtheorem{thm}{Theorem}[section]
\newtheorem{lem}[thm]{Lemma}
\newtheorem{cor}[thm]{Corollary}
\newtheorem{pro}[thm]{Proposition}

\theoremheaderfont{\scshape}

\newcommand{\ZM}{\mathbb{Z}}

\newcommand{\CM}{\mathbb{C}}

\newcommand{\fp}{\widetilde{f}^{(+)}}
\newcommand{\fm}{\widetilde{f}^{(-)}}
\newcommand{\lp}{\widetilde{\lambda}^{(+)}}
\newcommand{\lm}{\widetilde{\lambda}^{(-)}}

\title{{\Large {\bf A one-dimensional Hadamard walk with one defect}}}
\author{
{\small Takako Endo}\\
{\scriptsize  Department of Physics, Ochanomizu University}\\
{\scriptsize 2-1-1 Ohtsuka, Bunkyo, Tokyo, 112-0012, Japan}\\
{\scriptsize e-mail: g1170615@edu.cc.ocha.ac.jp}\\
{\small Norio Konno}\\
{\scriptsize Department of Applied Mathematics, 
Faculty of Engineering, 
Yokohama National University}\\
{\scriptsize 79-5 Hodogaya, Yokohama, 240-8501, Japan}\\
{\scriptsize e-mail: konno@ynu.ac.jp}\\
{\small Etsuo Segawa}\\
{\scriptsize Graduate School of Information Sciences, Tohoku University}\\
{\scriptsize 6-3-09 Aramaki Aza, Aoba, Sendai, Miyagi, 980-8579, Japan}\\
{\scriptsize e-mail: e.segawa@m.tohoku.ac.jp}\\
{\small Masato Takei}\\
{\scriptsize Department of Applied Mathematics, 
Faculty of Engineering, 
Yokohama National University}\\
{\scriptsize 79-5 Hodogaya, Yokohama, 240-8501, Japan}\\
{\scriptsize e-mail: takei@ynu.ac.jp}\\
}
\date{\empty}
\begin{document}
\maketitle

\par\noindent
\begin{small}
\par\noindent
{\bf Abstract}. We consider a one-dimensional space-inhomogeneous discrete time quantum walk. This model is the Hadamard walk with one defect at the origin which is different from the model introduced by Wojcik et al. \cite{WojcikEtAl2012}. We obtain a stationary measure of the model by solving the eigenvalue problem and an asymptotic behaviour of the return probability by the path counting approach. Moreover, we get the time-averaged limit measure using the space-time generating function method. The measure is symmetric for the origin and independent of the initial coin state at the starting point. So localization depends only on the parameter which determines the model. 
  
\footnote[0]{
{\it Abbr. title:} A one-dimensional Hadamard walk with one defect
}
\end{small}

\setcounter{equation}{0}

\section{Introduction \label{intro}}
The quantum walk (QW) has been investigated as a natural generalization of the classical random walk. This manuscript focuses on the discrete-time case. The QW on $\ZM$ was intensively studied by Ambainis et al. \cite{AmbainisEtAl2001}, where $\ZM$ is the set of integers. A number of non-classical properties of the QW have been shown, for example, ballistic spreading, anti-bellshaped limit density, localization. As for review and books on QWs, see Kempe \cite{Kempe2003}, Kendon \cite{Kendon2007}, Venegas-Andraca \cite{VAndraca2008, Venegas2013}, Konno \cite{Konno2008b}, Cantero et al. \cite{CanteroEtAl2013}, Manouchehri and Wang \cite{MW2013}. 

Wojcik et al. \cite{WojcikEtAl2012} introduced and investigated one-dimensional discrete time QW with one defect which is called ``the Wojcik model'' in this paper. Endo and Konno \cite{KW2013a} obtained a stationary measure for the Wojcik model solving the eigenvalue problem by the aid of the splitted generating function (SGF) method, which is consistent with  the result given in Wojcik et al. \cite{WojcikEtAl2012}. The SGF method is useful to find the stationary measure for the
QW with one defect in one dimension. Moreover, Endo and Konno \cite{KW2013b} got the time-averaged limit measure of the Wojcik model by several methods and found that the stationary measure is a special case of the time-averaged limit measure. The time-averaged limit measure is symmetric for the origin and localization depends heavily on the initial state and parameter (determines the model).

In this manuscript, we present another one-dimensional discrete time QW with one defect. For the one defect QW, we first obtain the stationary measure by the SGF method. From the path counting approach, we have a combinatorial expression of the return amplitude and its asymptotic behaviour. As a consequence, we get the time-averaged return probability which agrees with the result via the CGMV method \cite{cantero}. Furthermore, we present the time-averaged limit measure by the space-time generating function method. We should remark that the space-time generating function method does not allow us to get the stationary measure in our model.
 Like the corresponding measure of the Wojcik model, the measure is symmetric for the origin and localization depends on the model parameter $\xi$. However, we confirm that the time-averaged limit measure is independent of the initial coin state $\varphi$, so localization depends only on parameter $\xi$. Our model is suitable to consider the relation between the stationary measure and the time-averaged limit measure such as the Wojcik model.

The rest of the paper is organized as follows. In Sect. \ref{model}, we define our QW model. We obtain the stationary measure in Sect. \ref{stat}. The proofs of Proposition \ref{kyoui11D} and Lemma \ref{hodai11D} are devoted to Sects. \ref{meidai11D} and \ref{asitaka}, respectively. The asymptotic behaviour of the return probability amplitude is computed in Sect. \ref{konno-chap:qwOD-3}. Section \ref{seccgmv} deals with the result via the CGMV method. We give proofs of Proposition \ref{proppath} and Theorem \ref{gotennba} in Sects. \ref{secproppath} and \ref{proofthmpath}, respectively. In Sect. \ref{bokannsuu}, we explain the space-time generating function method. Section \ref{sectalm} gives the time-averaged limit measure by this method.

\section{Model \label{model}}
Let $\Psi_n (x)={}^T \! [\Psi^{L}_n(x),\Psi^{R}_n(x)]$ denote the amplitude of our model at time $n$ and position $x$, where $L$ and $R$ mean the left and right chirarities, respectively. Here $T$ stands for the transposed operator. First we prepare a sequences of $2 \times 2$ unitary matrices $\{ U_x : x \in \ZM \}$ given by 
\begin{align*}
U_x = 
\begin{bmatrix} 
a_{x} & b_{x} \\ 
c_{x} & d_{x}
\end{bmatrix},
\end{align*}
where $\ZM$ is the set of the integers. The time evolution of our model is determined by 
\begin{align*}
\Psi_{n+1} (x) = P_{x+1} \Psi_n (x+1) + Q_{x-1} \Psi_n (x-1) \quad (x \in \ZM),
\end{align*}
where
\begin{align*}
P_x = 
\begin{bmatrix} 
a_{x} & b_{x} \\ 
0 & 0 
\end{bmatrix}, 
\qquad 
Q_x = 
\begin{bmatrix} 
0 & 0 \\ 
c_{x} & d_{x}
\end{bmatrix}
\end{align*}
with $U_x = P_x + Q_x$. Then $P_x$ and $Q_x$ correspond to left and right movements, respectively. For our one-defect model, we define $U_x$ by
\begin{align}
U_{x}= 
\left\{ 
\begin{array}{ll}
\begin{bmatrix}
\cos \xi &  \sin \xi \\
\sin \xi & - \cos \xi 
\end{bmatrix} & \qquad (x=0), 
\\
\\
H & \qquad (x \in \ZM \setminus \{0\}),
\end{array} \right.
\end{align}
where $\xi \in (0, \pi/2)$. We can extend some cases to $\xi = 0$ or $\xi = \pi/2$. Here $H$ is the Hadamard matrix: 
\begin{align*}
H = 
\frac{1}{\sqrt{2}}
\begin{bmatrix}
1 &  1 \\
1 & -1 
\end{bmatrix}.
\end{align*}
From now on, we use notations $C= C(\xi) = \cos \xi$ and $S= S(\xi) = \sin \xi$. If $\xi = \pi/4$, then $U_x =H \> (x \in \ZM)$, i.e., this model becomes the well-known Hadamard walk. Therefore, our model can be considered as the Hadamard walk with one defect. We should remark that $\det (U_x) = -1$ for any $x \in \ZM$. Another Hadamard walk with one defect is the Wojcik model whose quantum coin $U_x$ at position $x$ is defined by
\begin{align}
U_{x}= 
\left\{ 
\begin{array}{ll}
\omega H & \qquad (x=0),  
\\
H & \qquad (x \in \ZM \setminus \{0\}),
\end{array} \right.
\end{align}
where $\omega = e^{2 \pi i \phi} \> (\phi \in (0,1))$. Then $\det (U_0) = - \omega^2$ is not equal to $\det (U_x) = \det (H) = -1$ for $x \not= 0$, so $\det (U_x)$ depends on the position for the Wojcik model. One of our motivations is that we want to know the influence of the position-dependence of $\det(U_{x})$ on the analysis.

\section{Stationary measure \label{stat}}
Let $\Psi(x)={}^T\![\Psi^{L}(x),\Psi^{R}(x)]$ denote the amplitude at position $x$. As in the case of the Wojcik model (see \cite{KW2013a}), we introduce the generating functions of $\Psi^{L}(x)$ and $\Psi^{R}(x)$, respectively, to get the stationary measure: 
\begin{align}
f^{j}_{+}(z) = \sum_{x=1}^{\infty} \Psi^{j}(x)z^{x}, \quad 
f^{j}_{-}(z) = \sum_{x=-1}^{-\infty} \Psi^{j}(x)z^{x} \qquad (j=L,R).
\end{align}
The quantum coin at the origin is different from that of the other position, so we consider both positive and negative parts. Then, the eigenvalue problem $U^{(s)}\Psi=\lambda\Psi$ is equivalent to
\begin{align*}
\lambda
\begin{bmatrix}
\Psi^{L}(x) \\ 
\Psi^{R}(x)
\end{bmatrix} 
=
\begin{bmatrix} 
a_{x+1} & b_{x+1} \\ 
0 & 0 
\end{bmatrix}
\begin{bmatrix} 
\Psi^{L}(x+1) \\ 
\Psi^{R}(x+1)
\end{bmatrix}
+
\begin{bmatrix} 
0 & 0 \\ 
c_{x-1} & d_{x-1}
\end{bmatrix} 
\begin{bmatrix} 
\Psi^{L}(x-1) \\ 
\Psi^{R}(x-1)
\end{bmatrix}
\quad (x \in \ZM).
\end{align*}
From the SGF method, we solve the eigenvalue problem $U^{(s)} \Psi = \lambda \Psi$ and obtain

\begin{pro}\label{kyoui11D}
Put $\alpha=\Psi^{L}(0)$ and $\beta=\Psi^{R}(0)$. Then, solution of the eigenvalue problem
\begin{align*}
U^{(s)} \Psi = \lambda \Psi
\end{align*}
is given in the following way. Here $\Psi= {}^T\![\cdots,\Psi^{L}(-1),\Psi^{R}(-1),\Psi^{L}(0),\Psi^{R}(0),\Psi^{L}(1),\Psi^{R}(1),\cdots]\in\mathbb{C}^{\infty}$, and $\lambda\in\mathbb{C}$ satisfies $|\lambda|=1$.
\par\noindent
(1) $\beta = - i \alpha$ case. We get
\begin{align*}
\lambda = \pm \dfrac{C+(\sqrt{2}-S)i}{\sqrt{3-2 \sqrt{2}S}}.
\end{align*}
Then, we have
\begin{align*}
\Psi^L(x) = 
\left\{
\begin{array}{ll}
\alpha \times \left( \pm \dfrac{i}{\sqrt{3-2 \sqrt{2}S}} \right)^x 
& (x \geq 1), \\
\alpha
& (x=0), \\
\left\{ \sqrt{2}C \alpha + (\sqrt{2}S - 1) \beta \right\} \times \left(\mp \dfrac{i}{\sqrt{3-2 \sqrt{2}S}} \right)^{-x} 
& (x\leq -1),
\end{array}\right.
\end{align*}
and
\begin{align*}
\Psi^R(x) = 
\left\{
\begin{array}{ll}
\left\{(1 - \sqrt{2}S) \alpha + \sqrt{2}C \beta \right\} \times \left( \pm \dfrac{i}{\sqrt{3-2 \sqrt{2}S}} \right)^x 
& (x \geq 1), \\
\beta
& (x=0), \\
\beta \times \left(\mp \dfrac{i}{\sqrt{3-2 \sqrt{2}S}} \right)^{-x} 
& (x\leq -1). 
\end{array}\right.
\end{align*}
\par\noindent
(2) $\beta = i \alpha$ case.  We get
\begin{align*}
\lambda = \pm \dfrac{C-(\sqrt{2}-S)i}{\sqrt{3-2 \sqrt{2}S}}.
\end{align*}
Then, we have 
\begin{align*}
\Psi^L(x) = 
\left\{
\begin{array}{ll}
\alpha \times \left( \mp \dfrac{i}{\sqrt{3-2 \sqrt{2}S}} \right)^x 
& (x \geq 1), \\
\alpha
& (x=0), \\
\left\{ \sqrt{2}C \alpha + (\sqrt{2}S - 1) \beta \right\} \times \left(\pm \dfrac{i}{\sqrt{3-2 \sqrt{2}S}} \right)^{-x} 
& (x\leq -1),
\end{array}\right.
\end{align*}
and
\begin{align*}
\Psi^R(x) = 
\left\{
\begin{array}{ll}
\left\{(1 - \sqrt{2}S) \alpha + \sqrt{2}C \beta \right\} \times \left( \mp \dfrac{i}{\sqrt{3-2 \sqrt{2}S}} \right)^x 
& (x \geq 1), \\
\beta
& (x=0), \\
\beta \times \left(\pm \dfrac{i}{\sqrt{3-2 \sqrt{2}S}} \right)^{-x} 
& (x\leq -1). 
\end{array}\right.
\end{align*}
\end{pro}

The proof of Proposition \ref{kyoui11D} is given in Sect. \ref{meidai11D}. Indeed, we confirm that $\Psi$ in this proposition is a solution of the eigenvalue problem $U^{(s)} \Psi = \lambda \Psi.$ The measure at position $x$ is defined by 
\begin{align*}
\mu(x)=\|\Psi(x)\|^{2}=|\Psi^{L}(x)|^{2}+|\Psi^{R}(x)|^{2}.
\end{align*}
So Proposition \ref{kyoui11D} gives the following stationary measure what we want. 
\begin{thm}
\label{sanosan} 
\begin{align*}
\mu(x)=
\left\{ \begin{array}{ll}
|c|^{2} & (x=0), \\
(2 - \sqrt{2}S) |c|^2 \left( \dfrac{1}{3-2 \sqrt{2}S} \right)^{|x|} & (x \neq 0),
\end{array} 
\right.
\end{align*}
where $\alpha = c/\sqrt{2}$ and $\beta = \pm ci/\sqrt{2}$ with $c \in \mathbb{C}$.
\end{thm}

This stationary measure is symmetric at the origin and has an exponential decay except for $S=1/\sqrt{2} \> (\xi = \pi/4)$, i.e., the Hadamard walk case. Moreover, when $0<S<1/\sqrt{2} \> (\xi \in (0,\pi/4))$, we have
\begin{align*}
\sum_{x \in \ZM} \mu(x) = \dfrac{3-2 \sqrt{2}S}{1-\sqrt{2}S} |c|^2.
\end{align*}
If we let $|c|=\sqrt{1- \sqrt{2}S}/\sqrt{3-2 \sqrt{2}S}$, then the following stationary {\it probability} measure is obtained: 
\begin{cor}
\label{yasumihayai}
If $0<S<1/\sqrt{2}$, that is, $\xi \in (0,\pi/4)$, then 
\begin{align*}
\mu(x)=
\left\{ \begin{array}{ll}
\dfrac{1- \sqrt{2}S}{3-2 \sqrt{2}S} & (x=0), 
\\
\\
\dfrac{(1- \sqrt{2}S)(2- \sqrt{2}S)}{3-2 \sqrt{2}S} \left( \dfrac{1}{3-2 \sqrt{2}S} \right)^{|x|} & (x \neq 0).
\end{array} 
\right.
\end{align*}
We should remark that the initial coin state satisfies $\alpha^2 + \beta^2 = 0$ in this case.
\end{cor}

\section{Proof of Proposition \ref{kyoui11D} \label{meidai11D}}
First we should remark that the eigenvalue problem we need to solve is equivalent to 
\begin{align}
\lambda\Psi(x)=P_{x+1}\Psi(x+1)+Q_{x-1}\Psi(x-1).
\label{bisyou1D}
\end{align}
where 
\begin{align*}
P_{x}
&=
\left\{
\begin{array}{ll}
\dfrac{1}{\sqrt{2}}
\begin{bmatrix} 1&1 \\ 0&0 \end{bmatrix} & \quad (x \in \ZM \setminus \{0\}), 
\\
\\
\begin{bmatrix} C&S \\ 0&0 \end{bmatrix} & \quad (x=0),
\end{array} \right.
\\
\\
Q_{x}
&=
\left\{
\begin{array}{ll}
\dfrac{1}{\sqrt{2}}\begin{bmatrix}0&0\\ 1&-1 \end{bmatrix} & \quad (x \in \ZM \setminus \{0\}), 
\\
\\
\begin{bmatrix}
0&0\\
S&-C 
\end{bmatrix}& \quad (x=0).
\end{array} \right.
\end{align*}
Here $C= \cos \xi$ and $S= \sin \xi$. Let $\Psi(x)={}^T\![\Psi^{L}(x), \Psi^{R}(x)]$. Equation \eqref{bisyou1D} implies 
\begin{enumerate}
\item $x\neq\pm1$ case. 
\begin{align}
\lambda\Psi^{L}(x)
&=\dfrac{1}{\sqrt{2}}\Psi^{L}(x+1)+\dfrac{1}{\sqrt{2}}\Psi^{R}(x+1),
\label{nogeson11Da}
\\
\lambda\Psi^{R}(x)
&=\dfrac{1}{\sqrt{2}}\Psi^{L}(x-1)-\dfrac{1}{\sqrt{2}}\Psi^{R}(x-1).
\label{nogeson11Db}
\end{align}
\item $x=1$ case.
\begin{align}
\lambda\Psi^{L}(1)
&=\dfrac{1}{\sqrt{2}}\Psi^{L}(2)+\dfrac{1}{\sqrt{2}}\Psi^{R}(2),
\label{nogeson21Da}
\\
\lambda\Psi^{R}(1)
&=S\Psi^{L}(0) - C \Psi^{R}(0).
\label{nogeson21Db}
\end{align}
\item $x=-1$ case.
\begin{align}
\lambda\Psi^{L}(-1)
&=C \Psi^{L}(0) + S \Psi^{R}(0),
\label{nogeson31Da}
\\
\lambda\Psi^{R}(-1)
&=\dfrac{1}{\sqrt{2}}\Psi^{L}(-2)-\dfrac{1}{\sqrt{2}}\Psi^{R}(-2).
\label{nogeson31Db}
\end{align}
\end{enumerate}
Then we have
\begin{lem}
\label{hodai11D}
\begin{align}
A f_{\pm}(z)= a_{\pm}(z).
\label{afa1D}
\end{align}
Here 
\begin{align*}
A
&= 
\begin{bmatrix}\lambda-\dfrac{1}{\sqrt{2}z}& -\dfrac{1}{\sqrt{2}z} \\ \\
-\dfrac{z}{\sqrt{2}}& \lambda+\dfrac{z}{\sqrt{2}}
\end{bmatrix}, 
\qquad 
f_{\pm}(z)=
\begin{bmatrix}
f^{L}_{\pm}(z)
\\
\\ 
f_{\pm}^{R}(z)
\end{bmatrix},
\\
a_{+}(z)
&=
\begin{bmatrix}
- \lambda \alpha \\ 
z (S \alpha - C \beta)
\end{bmatrix},
\qquad a_{-}(z)
=
\begin{bmatrix}
\dfrac{C \alpha + S \beta}{z}  \\
- \lambda \beta 
\end{bmatrix},
\end{align*}
where $\alpha=\Psi^{L}(0)$ and $\beta=\Psi^{R}(0)$.
\end{lem}
The proof is given in Sect. \ref{asitaka}. Noting 
\begin{align}
\det A=\dfrac{\lambda}{\sqrt{2}z}\left\{z^{2}-\sqrt{2}\left(\dfrac{1}{\lambda}-\lambda\right)z-1\right\},
\label{deta11D}
\end{align}
we define $\theta_{s}$ and $\theta_{l} \in\mathbb{C}$ satisfying 
\begin{align}
\det A=\dfrac{\lambda}{\sqrt{2}z}(z-\theta_{s})(z-\theta_{l}),
\label{deta21D}
\end{align}
with  $|\theta_{s}|\leq1\leq|\theta_{l}|$ .

By using Lemma \ref{hodai11D}, we will get $f_{\pm}^{L}(z)$ and $f_{\pm}^{R}(z)$. 
\begin{enumerate}
\item $f_{+}^{L}(z)$ case. Equation \eqref{afa1D} gives 
\begin{align*}
f^{L}_{+}(z)
&= \dfrac{1}{\det A}\left\{\left(\lambda+\dfrac{z}{\sqrt{2}}\right)(-\lambda\alpha)+\dfrac{S \alpha -C \beta}{\sqrt{2}} \right\}
\\
&= \dfrac{1}{\det A}\left(-\dfrac{\lambda \alpha}{\sqrt{2}} \right) \left\{ z + \dfrac{(\sqrt{2} \lambda^2 -S )\alpha + C \beta}{\lambda \alpha} \right\}.
\end{align*}
We put $\theta_{s} = - \dfrac{(\sqrt{2} \lambda^2 -S )\alpha + C \beta}{\lambda \alpha}$. Then we have
\begin{align*}
f^{L}_{+}(z)
&= -\dfrac{\alpha z}{z-\theta_{l}} 
=-\dfrac{\alpha z}{z+\dfrac{1}{\theta_{s}}}
=-\alpha\dfrac{z\theta_{s}}{z\theta_{s}+1}
\\
&= -\alpha(\theta_{s}z) 
\left\{ 1+(-\theta_{s}z)+(-\theta_{s}z)^{2}+(-\theta_{s}z)^{3}+\cdots \right\}
\end{align*}
Therefore, we see
\begin{align}
f^{L}_{+}(z)=\alpha\sum_{x=1}^{\infty}(-\theta_{s}z)^{x}.
\label{panda11D}
\end{align}
From Eq. \eqref{panda11D} and definition of $f_{+}^{L}(z)$, we have
\begin{align}
\Psi^{L}(x)=\alpha(-\theta_{s})^{x}\;\;\;(x=1,2,\cdots),
\end{align}
where
\begin{align}
\theta_{s}= - \dfrac{(\sqrt{2} \lambda^2 -S )\alpha + C \beta}{\lambda \alpha}.
\label{onene11D}
\end{align}
\item $f_{+}^{R}(z)$ case. From Eq. \eqref{afa1D}, we see
\begin{align}
f^{R}_{+}(z)
&=\dfrac{1}{\det A} \left[- \dfrac{\{ (1 - \sqrt{2} S) \alpha + \sqrt{2} C \beta \} \lambda}{\sqrt{2}} \right] 
\nonumber
\\
& \qquad \qquad \times 
\left[ z + \dfrac{S \alpha -  C \beta}{\{ (1 - \sqrt{2} S) \alpha + \sqrt{2} C \beta \} \lambda} \right].
\end{align}
Put $\theta_{s}= - \dfrac{S \alpha -  C \beta}{\{ (1 - \sqrt{2} S) \alpha + \sqrt{2} C \beta \} \lambda}$. Then we have
\begin{align}
f^{R}_{+}(z)
&=-\dfrac{\left\{ (1 - \sqrt{2} S) \alpha + \sqrt{2} C \beta \right\} z}{z-\theta_{l}}
\nonumber
\\
&=
\left\{ (1 - \sqrt{2} S) \alpha + \sqrt{2} C \beta \right\} \sum_{x=1}^{\infty}(-\theta_{s}z)^{x}.
\label{panda21D}
\end{align}
Combining Eq. \eqref{panda21D} with definition of $f_{+}^{R}(z)$ gives 
\begin{align}
\Psi^{R}(x)= \left\{ (1 - \sqrt{2} S) \alpha + \sqrt{2} C \beta \right\} \; (-\theta_{s})^{x}\;\;\;(x=1,2,\cdots),
\end{align}
where
\begin{align}
\theta_{s}=- \dfrac{S \alpha -  C \beta}{\{ (1 - \sqrt{2} S) \alpha + \sqrt{2} C \beta \} \lambda}.
\label{onene21D}
\end{align}
\item $f_{-}^{L}(z)$ case. Equation \eqref{afa1D} implies
\begin{align}
f^{L}_{-}(z)=\dfrac{C \alpha + S \beta}{\lambda(z-\theta_{l})(z-\theta_{s})}\left[ z+\dfrac{ \left\{  \sqrt{2} C \alpha + (\sqrt{2} S -1) \beta \right\} \lambda}{C \alpha + S \beta}\right].
\end{align}
Let $\theta_{l}=-\dfrac{\left\{  \sqrt{2} C \alpha + (\sqrt{2} S -1) \beta \right\} \lambda}{C \alpha + S \beta}$. Then we have
\begin{align*}
f^{L}_{-}(z)
&= \dfrac{C \alpha + S \beta}{\lambda}\times\dfrac{1}{z-\theta_{s}}
=\dfrac{C \alpha + S \beta}{\lambda}\times\dfrac{1}{\theta_{s}}\times\dfrac{\theta_{s}}{z}\dfrac{1}{1-\dfrac{\theta_{s}}{z}}
\\
&= \dfrac{C \alpha + S \beta}{\lambda}\times\dfrac{1}{\theta_{s}}\left\{\dfrac{\theta_{s}}{z}+\left(\dfrac{\theta_{s}}{z}\right)^{2}+\cdots\right\}.
\end{align*}
Thus we obtain
\begin{align}
f^{L}_{-}(z)
&=\dfrac{C \alpha + S \beta}{\lambda}\times\left(\dfrac{1}{\theta_{s}}\right)\sum^{-\infty}_{x=-1}(\theta^{-1}_{s}z)^{x}
\nonumber
\\
&= \left\{  \sqrt{2} C \alpha + (\sqrt{2} S -1) \beta \right\} \; \sum^{-\infty}_{x=-1}(\theta^{-1}_{s}z)^{x}
\label{panda31D}
\end{align}
Therefore, by Eq. \eqref{panda31D} and definition of $f^{L}_{-}(z)$, we have
\begin{align*}
\Psi^{L}(x)= \left\{  \sqrt{2} C \alpha + (\sqrt{2} S -1) \beta \right\} \; (\theta_{s})^{-x}\;\;\;(x=-1,-2,\cdots),
\end{align*}
where
\begin{align}
\theta_{s}=\dfrac{C \alpha + S \beta}{\left\{  \sqrt{2} C \alpha + (\sqrt{2} S -1) \beta \right\} \lambda}.
\label{onene31D}
\end{align}
\item $f_{-}^{R}(z)$ case. From Eq. \eqref{afa1D}, we see
\begin{align*}
f^{R}_{-}(z)
=\dfrac{1}{(z-\theta_{l})(z-\theta_{s})} \times \dfrac{C\alpha +(S- \sqrt{2} \lambda^2) \beta}{\lambda} \left\{ z+\dfrac{\lambda \beta}{C\alpha +(S- \sqrt{2} \lambda^2) \beta}\right\}.
\end{align*}
Put $\theta_{l}=- \dfrac{\lambda \beta}{C\alpha +(S- \sqrt{2} \lambda^2) \beta}$. Then we get
\begin{align*}
f^{R}_{-}(z)
&=\dfrac{C\alpha +(S- \sqrt{2} \lambda^2) \beta}{\lambda}\times\dfrac{1}{z-\theta_{s}}
\\
&=\dfrac{C\alpha +(S- \sqrt{2} \lambda^2) \beta}{\lambda}\times\dfrac{1}{\theta_{s}}\times\dfrac{\theta_{s}}{z}\times\dfrac{1}{1-\dfrac{\theta_{s}}{z}}.
\end{align*}
Thus, we have
\begin{align}
f^{R}_{-}(z)=\beta\sum^{-\infty}_{x=-1}(\theta_{s}^{-1}z)^{x}.
\label{panda41D}
\end{align}
Combining Eq. \eqref{panda41D} with definition of $f^{R}_{-}(z)$ implies
\begin{align*}
\Psi^{R}(x)=\beta(\theta_{s})^{-x}\;\;\;(x=-1,-2,\cdots), 
\end{align*}
where
\begin{align}
\theta_{s}=\dfrac{C\alpha +(S- \sqrt{2} \lambda^2) \beta}{\lambda \beta}.
\label{onene41D}
\end{align}
\end{enumerate} 
Therefore, we obtain 
\begin{align}
\Psi(x)=\left\{ \begin{array}{ll}
(-\theta_{s})^{x}
\begin{bmatrix} 
\alpha \\ 
(1- \sqrt{2}S) \alpha + \sqrt{2} C \beta
\end{bmatrix} &(x=1,2,\cdots), \\
\begin{bmatrix}
\alpha \\ \beta 
\end{bmatrix} &(x=0),\\
(\theta_{s})^{|x|}
\begin{bmatrix}
\sqrt{2} C \alpha + (\sqrt{2} S -1 ) \beta \\ 
\beta
\end{bmatrix} &(x=-1,-2,\cdots).
\end{array} \right.
\label{araisan1D}
\end{align}
\par\indent
Moreover, four expressions of $\theta_{s}$, that is, Eqs. \eqref{onene11D}, \eqref{onene21D}, \eqref{onene31D}, and \eqref{onene41D}, yield $\beta=i\alpha$ or $\beta=-i\alpha$. In fact, we have
\begin{align*}
\theta_{s}
&= - \dfrac{(\sqrt{2} \lambda^2 -S) \alpha + C \beta}{\lambda \alpha}= - \dfrac{S \alpha - C \beta}{\left\{ (1- \sqrt{2}S) \alpha + \sqrt{2} C \beta \right\} \lambda} 
\\
&= \dfrac{C \alpha + S \beta}{\left\{ \sqrt{2} C \alpha + (\sqrt{2} S -1 ) \beta \right\} \lambda} = \dfrac{C \alpha + (S - \sqrt{2} \lambda^2) \beta}{\lambda \beta}.
\end{align*}
The first and fourth expressions imply 
\begin{align*}
\alpha^2 + \beta^2 = 0.
\end{align*}
Thus we consider $\beta = \pm i \alpha$. For $\beta = - i \alpha$ case, the first and second expressions give
\begin{align*}
\lambda^2 = \frac{-(\sqrt{2}S-1)^2 - 2(S - \sqrt{2})C i}{3 - 2 \sqrt{2}S}.
\end{align*}
And the third expression implies 
\begin{align*}
\theta_s = \frac{\sqrt{2} -S - C i}{(3 - 2 \sqrt{2}S)\lambda}.
\end{align*}
Therefore, we have
\begin{align*}
\lambda 
= \pm \frac{C + (\sqrt{2} -S) i}{\sqrt{3 - 2 \sqrt{2}S}},
\qquad 
\theta_s 
= \mp \frac{i}{\sqrt{3 - 2 \sqrt{2}S}}.
\end{align*}
Similarly, for $\beta = i \alpha$ case, the first and second expressions give
\begin{align*}
\lambda^2 = \frac{-(\sqrt{2}S-1)^2 + 2(S - \sqrt{2})C i}{3 - 2 \sqrt{2}S}.
\end{align*}
And the third expression implies 
\begin{align*}
\theta_s = \frac{\sqrt{2} -S + C i}{(3 - 2 \sqrt{2}S)\lambda}.
\end{align*}
Thus, we get
\begin{align*}
\lambda 
= \pm \frac{C - (\sqrt{2} -S) i}{\sqrt{3 - 2 \sqrt{2}S}},
\qquad 
\theta_s 
= \pm \frac{i}{\sqrt{3 - 2 \sqrt{2}S}}.
\end{align*}
So we have the desired conclusion.

\section{Proof of Lemma \ref{hodai11D} \label{asitaka}}
By Eqs. \eqref{nogeson11Da} and \eqref{nogeson11Db}, we have
\begin{align*}
\lambda\sum^{\infty}_{x=2}\Psi^{L}(x)z^{x}
&=\frac{1}{\sqrt{2}}\sum^{\infty}_{x=2}\Psi^{L}(x+1)z^{x}+\frac{1}{\sqrt{2}}\sum^{\infty}_{x=2}\Psi^{R}(x+1)z^{x},
\\
\lambda\sum^{\infty}_{x=2}\Psi^{R}(x)z^{x}
&=\frac{1}{\sqrt{2}}\sum^{\infty}_{x=2}\Psi^{L}(x-1)z^{x}-\frac{1}{\sqrt{2}}\sum^{\infty}_{x=2}\Psi^{R}(x-1)z^{x}.
\end{align*}
From these equations, we see
\begin{align}
&\left(\lambda-\dfrac{1}{\sqrt{2}z}\right)f^{L}_{+}(z)-\dfrac{1}{\sqrt{2}z}f^{R}_{+}(z)
=\lambda z\Psi^{L}(1)-\dfrac{z}{\sqrt{2}}(\Psi^{L}(2)+\Psi^{R}(2))
\nonumber
\\
& \qquad \qquad \qquad \qquad \qquad \qquad \qquad \qquad \qquad 
-\dfrac{1}{\sqrt{2}}(\Psi^{L}(1)+\Psi^{R}(1)),
\label{nogeson41D}
\\
&
\left(\lambda+\dfrac{z}{\sqrt{2}}\right)f^{R}_{+}(z)-\dfrac{z}{\sqrt{2}}f^{L}_{+}(z)=\lambda z \Psi^{R}(1).
\label{nogeson51D}
\end{align}
From now on, we will express the right hand side of the above equations by using $\Psi^{L}(0)$ and $\Psi^{R}(0)$. First, we put $x=0$ in Eq. \eqref{nogeson11Da} and have
\begin{align*}
\lambda \Psi^{L}(0)= \frac{1}{\sqrt{2}}\Psi^{L}(1)+ \frac{1}{\sqrt{2}}\Psi^{R}(1).
\end{align*}
We substitute this equation and Eq. \eqref{nogeson21Da} into the right hand side of Eq. \eqref{nogeson41D}. Then we have 
\begin{align*}
\lambda z \Psi^{L}(1) - \lambda z \Psi^{L}(1) - \lambda \Psi^{L}(0) = - \lambda \Psi^{L}(0) = - \lambda \alpha.
\end{align*}
Similarly, Eq. \eqref{nogeson21Db} implies that the raight hand side of Eq. \eqref{nogeson51D} becomes
\begin{align*}
\lambda z \Psi^{R}(1) = z (S \Psi^{L}(0) - C \Psi^{R}(0)) = z (S \alpha - C \beta).
\end{align*}

Next, in a similar fashion, Eqs. \eqref{nogeson11Da} and \eqref{nogeson11Db} give
\begin{align*}
\lambda\sum^{-\infty}_{x=-2}\Psi^{L}(x)z^{x}
&=\frac{1}{\sqrt{2}}\sum^{-\infty}_{x=-2}\Psi^{L}(x+1)z^{x}+\frac{1}{\sqrt{2}}\sum^{-\infty}_{x=-2}\Psi^{R}(x+1)z^{x},
\\
\lambda\sum^{-\infty}_{x=-2}\Psi^{R}(x)z^{x}
&=\frac{1}{\sqrt{2}}\sum^{-\infty}_{x=-2}\Psi^{L}(x-1)z^{x}-\frac{1}{\sqrt{2}}\sum^{-\infty}_{x=-2}\Psi^{R}(x-1)z^{x}.
\end{align*}
From these equations, we have
\begin{align}
&\left(\lambda-\dfrac{1}{\sqrt{2}z}\right)f^{L}_{-}(z)-\dfrac{1}{\sqrt{2}z}f^{R}_{-}(z)= \frac{\lambda}{z} \Psi^{L}(-1),
\label{nogeson61D}
\\
&
\left(\lambda+\dfrac{z}{\sqrt{2}}\right)f^{R}_{-}(z)-\dfrac{z}{\sqrt{2}}f^{L}_{-}(z)
\nonumber
\\
&= \frac{\lambda}{z} \Psi^{R}(-1) -\frac{1}{\sqrt{2}z}(\Psi^{L}(-2)-\Psi^{R}(-2))-\dfrac{1}{\sqrt{2}}\Psi^{L}(-1)+\dfrac{1}{\sqrt{2}}\Psi^{R}(-1).
\label{nogeson71D}
\end{align}
Equation \eqref{nogeson31Da} implies that the right hand side of Eq. \eqref{nogeson61D} becomes
\begin{align*}
\frac{\lambda}{z} \Psi^{L}(-1) = \frac{C \Psi^{L}(0) + S \Psi^{R}(0)}{z} = \frac{C \alpha + S \beta}{z}.
\end{align*}
We put $x=0$ in Eq. \eqref{nogeson11Db} and get
\begin{align*}
\lambda \Psi^{R}(0)=\frac{1}{\sqrt{2}}\Psi^{L}(-1)-\frac{1}{\sqrt{2}}\Psi^{R}(-1).
\end{align*}
This equation and Eq. \eqref{nogeson31Db} imply that the right hand side of Eq. \eqref{nogeson71D} becomes
\begin{align*}
\frac{\lambda}{z} \Psi^{R}(-1) - \frac{\lambda}{z} \Psi^{R}(-1) - \lambda \Psi^{R}(0) = - \lambda \Psi^{R}(0) = - \lambda \beta.
\end{align*}
Therefore, the proof of Lemma \ref{hodai11D} is complete.

\section{Asymptotic behaviour \label{konno-chap:qwOD-3}}

Let the probability amplitude at time $2n$ be
\begin{align}
\Psi_{2n} (0) =  
\left[
\begin{array}{cc}
\Psi_{2n}^{L} (0) \\
\Psi_{2n}^{R} (0)
\end{array}
\right].
\end{align}
Remark that $\Psi_{2n+1} (0) = {}^T [\Psi_{2n+1}^{L} (0), \Psi_{2n+1}^{R} (0)] = {}^T [0,0].$ Then we have an expression of $\Psi_{2n} (0)$.
\begin{pro}
\label{proppath} We consider the QW stariting from the origin with the quantum bit $\varphi = {}^T [\alpha, \beta]$, where $\alpha, \beta \in \CM$ with $|\alpha|^2 + |\beta|^2 =1.$ Put \begin{align*}
r_n^{\ast} = 
\left\{
\begin{array}{cl}
\displaystyle{(-1)^{m-1} \> \frac{(2m-2)!}{2^{2m-1} (m-1)! m!}} & \quad \mbox{($n=4m-1 \>\> m \ge 1$),} \\
0 & \quad \mbox{($n \not= 4m-1, \> n \ge 2, \>\> m \ge 1$),} \\
-1 & \quad \mbox{($n =1$).}
\end{array}
\right.
\end{align*}
Then, we have for $n \ge 1$,
\begin{align*}
\Psi_{2n} (0) 
&= \sum_{k=1}^n \sum_{\scriptstyle (a_1, \ldots , a_k) \in (\ZM_{>})^k : \atop \scriptstyle a_1 + \cdots + a_k =n} \left( \prod_{j=1}^k r^{\ast}_{2 a_j-1} \right) 
\\
& \qquad \qquad \times \frac{1}{2}
\left[
\begin{array}{cc}
(\alpha - i \beta) \left( \dfrac{-S+Ci}{\sqrt{2}} \right)^k + (\alpha + i \beta) \left( - \dfrac{-S-Ci}{\sqrt{2}} \right)^k 
\\
i(\alpha - i \beta) \left( \dfrac{-S+Ci}{\sqrt{2}} \right)^k -i (\alpha + i \beta) \left( - \dfrac{-S-Ci}{\sqrt{2}} \right)^k 
\end{array}
\right].
\end{align*}

Here $\ZM_{>} = \{1,2, \ldots \}$ and 
\begin{align*}
\sum_{n=1}^{\infty} \> r_{n}^{\ast} z^n = \frac{-1 - z^2 + \sqrt{1 + z^4}}{z}.
\end{align*}
\end{pro}
The proof of Proposition \ref{proppath} appears in Sect. \ref{secproppath}. Let the return probability at the origin and at time $n$ be denoted by $r_n (0) = P(X_n=0)$. From this proposition, we obtain one of our main results, that is, the asymptotic behaviour of $\Psi_{2n} ^{L} (0)$ and $\Psi_{2n} ^{R} (0)$ as follows.

\begin{thm}
\label{gotennba}
\begin{align*}
\Psi_{2n} ^{L} (0)
& \sim \frac{2(1- \sqrt{2}S)}{3-2 \sqrt{2}S} \> \left\{ \cos (n \theta_0) \> \alpha - \sin (n \theta_0) \> \beta \right\} \times I_{[0, \pi/4)}(\xi),
\\
\Psi_{2n} ^{R} (0)
& \sim \frac{2(1- \sqrt{2}S)}{3-2 \sqrt{2}S} \> \left\{ \cos (n \theta_0) \> \beta + \sin (n \theta_0) \> \alpha \right\} \times I_{[0, \pi/4)}(\xi),
\end{align*}
where $I_A (x) =1 \> (x \in A), \> =0 \> (x \not\in A)$, and
\begin{align*}
\cos \theta_0 = - \frac{(1- \sqrt{2}S)^2}{3-2 \sqrt{2}S}, \quad \sin \theta_0 = \frac{2 (\sqrt{2}-S)C}{3-2 \sqrt{2}S}.
\end{align*}
Here, $f(n) \sim g(n)$ means $f(n)/g(n) \to 1 \>\> (n \to \infty)$.
\end{thm}
The proof of Theorem \ref{gotennba} is given in Sect. \ref{proofthmpath}. By this theorem, we have
\begin{align*}
\left| \Psi_{2n} ^{L} (0) \right|^2 
& \sim \frac{4(1- \sqrt{2}S)^2}{(3-2 \sqrt{2}S)^2} \> \left\{ \cos^2 (n \theta_0) \> |\alpha|^2 + \sin^2 (n \theta_0) \> |\beta|^2 \right.
\\
& \qquad \qquad \left. - \cos (n \theta_0) \sin (n \theta_0) (\alpha \overline{\beta} + \overline{\alpha} \beta ) \right\} \times I_{[0, \pi/4)}(\xi),
\\
\left| \Psi_{2n} ^{R} (0) \right|^2 
& \sim \frac{4(1- \sqrt{2}S)^2}{(3-2 \sqrt{2}S)^2} \> \left\{ \cos^2 (n \theta_0) \> |\beta|^2 + \sin^2 (n \theta_0) \> |\alpha|^2 \right.
\\
& \qquad \qquad \left. + \cos (n \theta_0) \sin (n \theta_0) (\alpha \overline{\beta} + \overline{\alpha} \beta ) \right\} \times I_{[0, \pi/4)}(\xi).
\end{align*}
The definition of $r_{2n} (0)$ implies 
\begin{align*}
r_{2n} (0) = |\Psi_{2n} ^{L} (0)|^2 + |\Psi_{2n} ^{R} (0)|^2  \sim \frac{4(1- \sqrt{2}S)^2}{(3-2 \sqrt{2}S)^2} \>  \times I_{[0, \pi/4)}(\xi).
\end{align*}
Thus we have the limit of $r_{2n} (0)$. Moreover, noting $r_{2n+1} (0)=0$, we get the time-averaged limit measure at the origin, $\overline{\mu}_{\infty} (0)$, as follows.
\begin{cor}
\label{corpath}
\begin{align*}
\lim_{n \to \infty} \> r_{2n} (0) 
&= \frac{4(1 - \sqrt{2}S)^2}{(3 - 2 \sqrt{2}S)^2} \times I_{(0,\pi/4)}(\xi), 
\\
\overline{\mu}_{\infty} (0) 
&= \frac{2(1 - \sqrt{2}S)^2}{(3 - 2 \sqrt{2}S)^2} \times I_{(0,\pi/4)}(\xi).
\end{align*}
\end{cor}
If $\xi \in [0, \pi/4)$, then localization occurs. If $\xi \in [\pi/4, \pi/2)$, then localization does not occur. Remark that when $\xi = \pi/4$, the model becomes the Hadamard walk.

\section{Result via the CGMV method \label{seccgmv}}
We can derive the time-averaged limit measure at the origin $\overline{\mu}_{\infty}(0)$ also from the CGMV method \cite{cantero}. From now on, we use the same notations as in Ref. \cite{cantero}. Applying the CGMV method to our model, we have
\begin{align*}
a=\dfrac{i}{\sqrt{2}}, \quad b=iS,\quad \omega = 1,\quad \zeta_{\pm}(b)=\pm C +Si.
\end{align*}
As the conditions ${\mathcal M}_{\pm}$, we see that the following same inequality holds.
\begin{align*}
0<\xi<\frac{\pi}{4}.
\end{align*}
Moreover, we have
\begin{align*}
\sigma_{1}=0,\;\sigma_{2}=\pi,\quad \sigma=\sigma_{1}+\sigma_{2}=\pi,\quad \theta=\dfrac{\sigma}{2}=\dfrac{\pi}{2},\\
\tau_{1}=0, \quad \tau_{2}= \pi,\quad \tau=\tau_{1}+\tau_{2}= \pi.
\end{align*}
According to the CGMV method, we get
\begin{align}
\lim_{n\to\infty}P^{(0)}_{\alpha,\beta}(2n)=\dfrac{1}{2}\left(1-\dfrac{\rho_{a}^{2}}{|\zeta_{\pm}(b)-a|^{2}}\right)^{2}
\left\{
1\mp\dfrac{(|\hat{\alpha}|^{2}-|\hat{\beta}|^{2})\Re b+2\rho_{b}\Re(\overline{\omega\hat{\alpha}}\hat{\beta})}
{\sqrt{1-\Im^{2}b}}
\right\},
\label{300}
\end{align}
where $P^{(0)}_{\alpha,\beta}(2n)$ is the probability that the walker return to the origin at time $2n$ with the initial qubit $\varphi={}^T\![\alpha,\beta]$, where $\alpha,\beta\in\mathbb{C}$ and $|\alpha|^{2}+|\beta|^{2}=1$. Here,
\begin{align*}
&\rho_{a}=\dfrac{1}{\sqrt{2}},\quad|\zeta_{\pm}(b)-a|^{2}=\dfrac{3}{2}- \sqrt{2}S, 
\nonumber
\\
&\Im b=S,\quad \Re b=0,\quad \rho_{b}=C,\quad \hat{\alpha}=\hat{\lambda}^{(1)}_{0}\alpha=\alpha,
\nonumber\\
&\hat{\beta} =\hat{\lambda}^{(2)}_{1}\beta = e^{i\left((\sigma_{2}-\sigma_{1})/2+\tau_{2}-\sigma_{2}\right)}\beta= i \beta.
\end{align*}
Therefore, Eq. (\ref{300}) becomes
\begin{align*}
\lim_{n\to\infty}P^{(0)}_{\alpha,\beta}(2n)=\left\{\begin{array}{ll}
\dfrac{2(1 - \sqrt{2}S)^2}{(3 - 2 \sqrt{2}S)^2}|\alpha-i\beta|^{2}& ({\mathcal M}_{+}, i.e., \xi \in (0,\pi/4)),\\
& \\
\dfrac{2(1 - \sqrt{2}S)^2}{(3 - 2 \sqrt{2}S)^2}|\alpha+i\beta|^{2}& ({\mathcal M}_{-}, i.e., \xi \in (0,\pi/4)).\\
\end{array} \right.
\end{align*} 
Thus, we obtain 
\begin{align*}
\lim_{n \to \infty} \> r_{2n} (0) 
&= \frac{4(1 - \sqrt{2}S)^2}{(3 - 2 \sqrt{2}S)^2} \times I_{[0,\pi/4)}(\xi), 
\\
\overline{\mu}_{\infty} (0) 
&= \frac{2(1 - \sqrt{2}S)^2}{(3 - 2 \sqrt{2}S)^2} \times I_{[0,\pi/4)}(\xi).
\end{align*}
These agrees with our result, Corollary \ref{corpath}.

\section{Proof of Proposition \ref{proppath} \label{secproppath}}
In this section, we prove Proposition \ref{proppath}. To do so, we consdier the Hadamard walk starting from $m \> (\ge 1)$ on  $\ZM_{\ge} = \{ 0, 1, 2, \ldots \}$ with absorbing boundary at the origin. The dynamics depends only on $\{ U_x \equiv H : x \ge 1\}$. Thus, we should remark that for any $x \ge 1$, 
\begin{align*}
P_x = P = \frac{1}{\sqrt{2}}
\left[
\begin{array}{cc}
1 & 1 \\
0 & 0
\end{array}
\right],
\quad
Q_x = Q = \frac{1}{\sqrt{2}}
\left[
\begin{array}{cc}
0 & 0 \\
1 & -1
\end{array}
\right] \qquad (x \ge 1).
\end{align*}
Next, we let $\Xi^{(\infty,m)} _n$ be the sum of weights over all paths starting the origin, moving on $\mathbb{Z}_{\geq}$,
 and returning to the origin for the first time at time $n$. For example, 
\begin{align*} 
\Xi^{(\infty,1)} _5  = P^2 Q P Q + P^3 Q^2.
\end{align*} 
Here, we introduce $R$ and $S$ as follows:
\begin{align*}
R =
\frac{1}{\sqrt{2}}
\left[
\begin{array}{cc}
1 & -1 \\
0 & 0 
\end{array}
\right], 
\quad
S=
\frac{1}{\sqrt{2}}
\left[
\begin{array}{cc}
0 & 0 \\
1 & 1 
\end{array}
\right].
\end{align*} 
Then $P, Q, R, S$ are an orthonormal basis of the vector space of complex $2 \times 2$ matricies with respect to the trace inner product $\langle A | B \rangle = $ tr$(A^{\ast}B)$, where $\ast$ means the adjoint operator. Thus, $\Xi^{(\infty,m)} _n$ can be uniquely expressed as 
\begin{align*} 
\Xi^{(\infty,m)} _n = p^{(\infty,m)} _n P + q^{(\infty,m)} _n Q + r^{(\infty,m)} _n R + s^{(\infty,m)} _n S.
\end{align*}
Noting the definition of $\Xi^{(\infty,m)} _n$, we see that for $m \ge 1$, 
\begin{align*} 
\Xi^{(\infty,m)} _n = \Xi^{(\infty,m-1)} _{n-1} P + \Xi^{(\infty,m+1)} _{n-1} Q.\end{align*}
By using this, we have
\begin{align*}
p^{(\infty,m)} _n 
&= \frac{1}{\sqrt{2}} \> p^{(\infty,m-1)} _{n-1} + \frac{1}{\sqrt{2}} \> r^{(\infty,m-1)} _{n-1}, 
\\  
q^{(\infty,m)} _n 
&= - \frac{1}{\sqrt{2}} \> q^{(\infty,m+1)} _{n-1} + \frac{1}{\sqrt{2}} \> s^{(\infty,m+1)} _{n-1}, 
\\ 
r^{(\infty,m)} _n 
&= \frac{1}{\sqrt{2}} \> p^{(\infty,m+1)} _{n-1} - \frac{1}{\sqrt{2}} \> r^{(\infty,m+1)} _{n-1}, 
\\ 
s^{(\infty,m)} _n 
&= \frac{1}{\sqrt{2}} \> q^{(\infty,m-1)} _{n-1} + \frac{1}{\sqrt{2}} \> s^{(\infty,m-1)} _{n-1}. 
\end{align*}
Moreover, the definition of $\Xi^{(\infty,m)} _n$ implies that the possible paths can be expressed as the following two types, $P \ldots P$ and $P \ldots Q$, since the last weight is $P$. Then, we have $q^{(\infty,m)} _n=s^{(\infty,m)} _n=0 \> (n \ge 1)$. In order to compute $p^{(\infty,m)} _n$ and $r^{(\infty,m)} _n$, we introduce the following generating functions: 
\begin{align*}
p^{(\infty,m)}  (z) = \sum_{n=1} ^{\infty} p^{(\infty,m)} _n z^n, \quad r^{(\infty,m)}  (z) = \sum_{n=1} ^{\infty} r^{(\infty,m)} _n z^n.
\end{align*}
Therefore, we have
\begin{align*}
p^{(\infty,m)} (z) &= \frac{z}{\sqrt{2}} \> p^{(\infty,m-1)} (z) + \frac{z}{\sqrt{2}} \> r^{(\infty,m-1)} (z), \\
r^{(\infty,m)} (z) &= \frac{z}{\sqrt{2}} \> p^{(\infty,m+1)} (z) - \frac{z}{\sqrt{2}} \> r^{(\infty,m+1)} (z). 
\end{align*} 
From these equations, we see that $p^{(\infty,m)} (z)$ and $r^{(\infty,m)} (z)$  satisfy    
\begin{align*} 
p^{(\infty,m+2)} (z) + \sqrt{2} \> \left( {1 \over z} - z \right) p^{(\infty,m+1)} (z) - p^{(\infty,m)} (z) &= 0,
\\
r^{(\infty,m+2)} (z) + \sqrt{2} \> \left( {1 \over z} - z \right) r^{(\infty,m+1)} (z) - r^{(\infty,m)} (z) &= 0.
\end{align*} 
Thus, the characteristic equation has the following two roots: 
\begin{align*}
\lambda_{\pm} =\frac{-1 + z^2 \pm \sqrt{1+z^4}}{\sqrt{2}z}.
\end{align*}

Next, the definition of $\Xi^{(\infty,1)} _n$ implies $p^{(\infty,1)}_n = 0 \> (n \ge 2)$ and $p_1 ^{(\infty,1)} =1$.  Thus, we have $p^{(\infty,1)}(z)=z.$ Moreover,
the definition of $\Psi_n^{(\infty)}$ gives $\lim_{m \to \infty}
p^{(\infty,m)}(z)=0.$
Similarly, we have $\lim_{m \to \infty} r^{(\infty,m)}(z)=0.$ Combining
$p^{(\infty,1)}(z)=z$
with  $\lim_{m \to \infty} p^{(\infty,m)}(z)=\lim_{m \to \infty}r^{(\infty,m)}(z)=0$,
we get　
\begin{align*}
p^{(\infty,m)} (z) = z \lambda_ +^{m-1}, \quad r^{(\infty,m)} (z) = \frac{-1+\sqrt{1+z^4}}{z} \lambda_+^{m-1}.
\end{align*}
Therefore, for any $m=1$, we have
\begin{align*}
r^{(\infty,1)} (z) = \frac{-1+\sqrt{1+ z^4}}{z}.
\end{align*}

In a similar fashion, we consdier the hadamard walk starting from $m (\le -1)$ on $\ZM_{\le} = \{ 0, -1, -2, \ldots \}$ with absorbing boundary at the origin. 
\begin{align*}
q^{(-\infty,m)} (z) = z \lambda_- ^{m+1}, \quad s^{(-\infty,m)} (z) = \frac{1-\sqrt{1+ z^4}}{z} \lambda_-^{m+1}.
\end{align*}
Thus, for any $m=-1$, 
\begin{align*}
s^{(- \infty, -1)} (z) = \frac{1 - \sqrt{1 + z^4}}{z}.
\end{align*} 
It is easily checked that for any $n \ge 1$, $r_{n}^{(\infty,1)} + s_{n}^{(-\infty,-1)} = 0$. Here we put $\Xi_n^{+} = \Xi_{n-1}^{(\infty, 1)} Q_0$ and $\Xi_n^{-} = \Xi_{n-1}^{(-\infty, -1)} P_0$, where
\begin{align*}
P_0 = 
\left[
\begin{array}{cc}
C & S \\
0 & 0 
\end{array}
\right], 
\quad
Q_0 = 
\left[
\begin{array}{cc}
0 & 0 \\
S & -C
\end{array}
\right].
\end{align*}
That is, $\Xi_n^{+}$ (resp. $\Xi_n^{-}$) is the sum of all paths with the weights that the quantum walker restricted in region $\ZM_{\ge}$ (resp. $\ZM_{\le}$) reaches the origin for the first time at time $n$. Therefore, we have
\begin{lem}
\label{konno-chap:qwODlem1}
(i) When $n \ge 4$ and $n$ is even, 
\begin{align*}
\Xi_n^{+} 
&= r^{(\infty,1)}_{n-1} \> R Q_0 
= \frac{r^{(\infty,1)}_{n-1}}{\sqrt{2}} 
\left[
\begin{array}{cc}
-S & C \\
0 & 0
\end{array}
\right],
\\
\Xi_n^{-} 
&= s^{(-\infty,-1)}_{n-1} \> S P_0 
= \frac{s^{(-\infty,-1)}_{n-1}}{\sqrt{2}} 
\left[
\begin{array}{cc}
0 & 0 \\
C & S
\end{array}
\right],
\end{align*}
where
\begin{align*}
\sum_{n=1}^{\infty} \> r_{n}^{(\infty,1)} z^n = \frac{-1 + \sqrt{1 + z^4}}{z}, \quad \sum_{n=1}^{\infty} \> s_{n}^{(-\infty,-1)} z^n = \frac{1 - \sqrt{1 + z^4}}{z}.
\end{align*}
(ii) 
\begin{align*}
\Xi_2^{+} 
= P Q_0 
= \frac{-1}{\sqrt{2}} 
\left[
\begin{array}{cc}
-S & C \\
0 & 0
\end{array}
\right], \qquad
\Xi_2^{-} 
= Q P_0 
= \frac{1}{\sqrt{2}} 
\left[
\begin{array}{cc}
0 & 0 \\
C & S
\end{array}
\right].
\end{align*}
(iii) When $n$ is odd, 
\begin{align*}
\Xi_n^{+} = \Xi_n^{-} 
= 
\left[
\begin{array}{cc}
0 & 0 \\
0 & 0
\end{array}
\right].
\end{align*}
\end{lem}

Here we put $\Xi_n^{\ast} = \Xi_n^{+} + \Xi_n^{-}$. From this lemma and $s_{n}^{(-\infty,-1)} = - r_{n}^{(\infty,1)} \> (n \ge 1)$, we have
\begin{align*}
\Xi^{\ast}_n
=
\frac{r^{\ast}_{n-1}}{\sqrt{2}} \> 
\left[
\begin{array}{cc}
-S & C \\
-C & -S
\end{array}
\right],
\end{align*}
where
\begin{align*}
r_n^{\ast} = 
\left\{
\begin{array}{cl}
\displaystyle{(-1)^{m-1} \> \frac{(2m-2)!}{2^{2m-1} (m-1)! m!}} & \quad \mbox{($n=4m-1 \>\> m \ge 1$),} \\
0 & \quad \mbox{($n \not= 4m-1, \> n \ge 2, \>\> m \ge 1$),} \\
-1 & \quad \mbox{($n =1$).}
\end{array}
\right.
\end{align*}
Indeed, we have
\begin{align*}
r_1^{\ast} 
&= -1, \> r_2^{\ast} = 0, \> r_3^{\ast} = 1/2,  \> r_4^{\ast} = r_5^{\ast} = r_6^{\ast} = 0, \> \\
r_7^{\ast} 
&= -1/8, \> r_8^{\ast} = r_9^{\ast} = r_{10}^{\ast} = 0, \ldots.   
\end{align*}
Then, the generating function of $r_{n}^{\ast}$ is given by
\begin{align*}
\sum_{n=1}^{\infty} \> r_{n}^{\ast} z^n = \frac{-1 - z^2 + \sqrt{1 + z^4}}{z}.
\end{align*}
From the definition of $\Xi^{\ast}_{n}$, we have
\begin{align*}
\Psi_{2n} (0) 
= \sum_{k=1}^n \sum_{\scriptstyle (a_1, \ldots , a_k) \in (\ZM_{>})^k : \atop \scriptstyle a_1 + \cdots + a_k =n} \left( \prod_{j=1}^k \Xi^{\ast}_{2 a_j} \right) \> \varphi,
\end{align*}
where $\ZM_{>} = \{1,2, \ldots \}$. Moreover, 
\begin{align*}
&\left( \frac{1}{\sqrt{2}} \right)^k
\>
\left[
\begin{array}{cc}
-S & C \\
-C & S
\end{array}
\right]^k 
\> 
\left[
\begin{array}{cc}
\alpha \\
\beta
\end{array}
\right]
\\
& \qquad = 
\frac{1}{2} 
\>
\left( \frac{1}{\sqrt{2}} \right)^k
\>
\left[
\begin{array}{cc}
1 & 1 \\
i & -i
\end{array}
\right]
\>
\left[
\begin{array}{cc}
(-S+Ci)^k & 0 \\
0 & (-S-Ci)^k
\end{array}
\right] 
\> 
\left[
\begin{array}{cc}
1 & -i \\
1 & i
\end{array}
\right]
\> 
\left[
\begin{array}{cc}
\alpha \\
\beta
\end{array}
\right]
\\
& \qquad =
\frac{1}{2}
\left[
\begin{array}{cc}
(\alpha - i \beta) \left( \dfrac{-S+Ci}{\sqrt{2}} \right)^k + (\alpha + i \beta) \left(\dfrac{-S-Ci}{\sqrt{2}} \right)^k 
\\
i(\alpha - i \beta) \left( \dfrac{-S+Ci}{\sqrt{2}} \right)^k -i (\alpha + i \beta) \left(\dfrac{-S-Ci}{\sqrt{2}} \right)^k 
\end{array}
\right].
\end{align*}
From Lemma \ref{konno-chap:qwODlem1} (iii), we have 
\begin{align*}
\left( \prod_{j=1}^k \Xi^{\ast}_{2 a_j} \right) \> \varphi
= 
\left( \prod_{j=1}^k r^{\ast}_{2 a_j-1} \right) \> 
\left( \frac{1}{\sqrt{2}} \right)^k
\>
\left[
\begin{array}{cc}
-S & C \\
-C & -S
\end{array}
\right]^k 
\> 
\left[
\begin{array}{cc}
\alpha \\
\beta
\end{array}
\right].
\end{align*}
Therefore, we have the desired conclusion. 

\section{Proof of Theorem \ref{gotennba} \label{proofthmpath}}
By using Proposition \ref{proppath}, we compute the generating function of $\Psi_{n} ^{L} (0)$. We put $x_n = r^{\ast}_{2n-1}$ and $u=(-S+Ci)/\sqrt{2}, \>\> \widetilde{u}=(-S-Ci)/\sqrt{2}$. Then we have
\begin{align*}
\sum_{n=1}^{\infty} \Psi_{2n} ^{L} (0) z^{2n} 
& = \frac{\alpha - i \beta}{2} \> 
\sum_{n=1}^{\infty} \Biggl\{ \sum_{k=1}^n \sum_{\scriptstyle (a_1, \ldots , a_k) \in (\ZM_{>})^k : \atop \scriptstyle a_1 + \cdots + a_k =n} \left( \prod_{j=1}^k x_{a_j} \right) u^k \Biggr\} z^{2n}
\\
&+ 
\frac{\alpha + i \beta}{2} \> 
\sum_{n=1}^{\infty} \Biggl\{ \sum_{k=1}^n \sum_{\scriptstyle (a_1, \ldots , a_k) \in (\ZM_{>})^k : \atop \scriptstyle a_1 + \cdots + a_k =n} \left( \prod_{j=1}^k x_{a_j} \right) \widetilde{u}^k \Biggr\} z^{2n}
\\
&=
\frac{\alpha - i \beta}{2} \> 
\sum_{k=1}^{\infty} \Biggl\{ \sum_{n=k}^{\infty} \sum_{\scriptstyle (a_1, \ldots , a_k) \in (\ZM_{>})^k : \atop \scriptstyle a_1 + \cdots + a_k =n} \left( \prod_{j=1}^k x_{a_j} \right)  z^{2n} \Biggr\} u^k  
\\
&+ 
\frac{\alpha + i \beta}{2} \> 
\sum_{k=1}^{\infty} \Biggl\{ \sum_{n=k}^{\infty} \sum_{\scriptstyle (a_1, \ldots , a_k) \in (\ZM_{>})^k : \atop \scriptstyle a_1 + \cdots + a_k =n} \left( \prod_{j=1}^k x_{a_j} \right) z^{2n} \Biggr\} \widetilde{u}^k.
\end{align*}
The first equality comes from Proposition \ref{proppath}. Thus, we get
\begin{align*}
\sum_{n=1}^{\infty} \Psi_{2n} ^{L} (0) z^{2n}  
&=  
\frac{\alpha - i \beta}{2} \> 
\sum_{k=1}^{\infty} \left\{ (-1 - z^2 + \sqrt{1 + z^4}) \> u \right\}^k 
\\
&+
\frac{\alpha + i \beta}{2} \> 
\sum_{k=1}^{\infty} \left\{ (-1 - z^2 + \sqrt{1 + z^4}) \> \widetilde{u} \right\}^k 
\\
&= \frac{\alpha - i \beta}{2} \> \frac{ (-1 - z^2 + \sqrt{1 + z^4}) u}{1 - (-1 - z^2 + \sqrt{1 + z^4}) u}
\\
&+ \frac{\alpha + i \beta}{2} \> \frac{ (-1 - z^2 + \sqrt{1 + z^4}) \widetilde{u}}{1 - (-1 - z^2 + \sqrt{1 + z^4}) \widetilde{u}}.
\end{align*}
As for the first equality, we see that for any $k \ge 1$, 
\begin{align*}
\sum_{n=k}^{\infty} \sum_{\scriptstyle (a_1, \ldots , a_k) \in (\ZM_{>})^k : \atop \scriptstyle a_1 + \cdots + a_k =n} \left( \prod_{j=1}^k x_{a_j} \right)  z^{2n} = (-1 - z^2 + \sqrt{1 + z^4})^k.
\end{align*}
Noting the initial state $\Psi_{0} ^{L} (0) = \alpha$, we have
\begin{align}
\sum_{n=0}^{\infty} \Psi_{2n} ^{L} (0) w^{n} 
&= \frac{\alpha - i \beta}{2} \>  \frac{1}{1 - Z u} +  \frac{\alpha + i \beta}{2} \> \frac{1}{1 - Z \widetilde{u}}
\nonumber
\\
&= \frac{\sqrt{2}\{ (\sqrt{2}+SZ) \alpha + C Z \beta \}}{2 + 2 \sqrt{2} S Z + Z^2},
\label{harugakite}
\end{align}
where $Z = -1 - w + \sqrt{1 + w^2}$. As for the generating function of $\Psi_{n} ^{R} (0)$, noting that $\Psi_{0} ^{R} (0) = \beta$, we similarly obtain
\begin{align*}
\sum_{n=0}^{\infty} \Psi_{n} ^{R} (0) z^{n} 
&= i \left( \frac{\alpha - i \beta}{2} \>  \frac{1}{1 - Z u} -  \frac{\alpha + i \beta}{2} \> \frac{1}{1 - Z \widetilde{u}} \right)
\\
&=  \frac{\sqrt{2}\{ (\sqrt{2}+SZ) \beta - C Z \alpha \}}{2 + 2 \sqrt{2} S Z + Z^2}.
\end{align*}
Thus, if we have the asymptotic behaviour of $\Psi_{2n} ^{L} (0)$ as $n \to \infty$, then we also have that of $\Psi_{2n} ^{R} (0)$ by $\alpha \to \beta$ and $\beta \to - \alpha$. So it is sufficient to consider the case of $\Psi_{2n} ^{L} (0)$. From now on, we will study an asymptotic behaviour of $\Psi_{2n} ^{L} (0)$. From Eq. \eqref{harugakite}, we get
\begin{lem}
We put
\begin{align*}
\sum_{n=0}^{\infty} \Psi_{2n} ^{L} (0) w^{n} 
= \{ A^{\alpha} _1 (w) + A^{\alpha} _2 (w) \} \> \alpha + \{ A^{\beta} _1 (w) + A^{\beta} _2 (w) \} \> \beta,
\end{align*}
where
\begin{align*}
A^{\alpha}_1 (w) 
&= \dfrac{2 \sqrt{2}-3S + \sqrt{2}(1-\sqrt{2}S)^2 w + (\sqrt{2} - S)w^2}{B(w)}
\\
A^{\alpha}_2 (w) 
&=\dfrac{(\sqrt{2}-S)(1+w)\sqrt{1+w^2}}{B(w)},
\\
A^{\beta}_1 (w) 
&= \dfrac{-C \left\{1 + 2 (1- \sqrt{2}S)w +w^2 \right\}}{B(w)},
\\
A^{\beta}_2 (w) 
&= \dfrac{C(1-w)\sqrt{1+w^2}}{B(w)},
\\
B (w) 
&=\sqrt{2}(3-2 \sqrt{2}S) \left\{ 1+ \dfrac{2(1-\sqrt{2}S)^2}{3-2 \sqrt{2}S}w +w^2\right\}.
\end{align*}
\end{lem}

First, we consider the following case:
\begin{align*}
A^{\alpha}_1 (w) 
=\dfrac{2 \sqrt{2}-3S + \sqrt{2}(1-\sqrt{2}S)^2 w + (\sqrt{2} - S)w^2}{\sqrt{2}(3-2 \sqrt{2}S) \left\{ 1+ \dfrac{2(1-\sqrt{2}S)^2}{3-2 \sqrt{2}S}w +w^2\right\} }.
\end{align*}
The two roots of 
\begin{align}
 1+ \dfrac{2(1-\sqrt{2}S)^2}{3-2 \sqrt{2}S}w +w^2 =0
\label{gotejobs1}
\end{align}
are denoted by $\gamma=e^{i \theta_0}$ and $\overline{\gamma}=e^{-i \theta_0}$ with 
\begin{align*}
\cos \theta_0 = - \frac{2(1-\sqrt{2}S)^2}{3-2 \sqrt{2}S} \quad (<0).
\end{align*}
Thus, we have
\begin{align*}
A^{\alpha}_1 (w) = \frac{1-\sqrt{2}S}{2(3-2 \sqrt{2}S)} \times \frac{1-w^2}{(w - \gamma) (w - \overline{\gamma})}.
\end{align*}
In general, when we have
\begin{align*}
f(z) = \sum_{n=0}^{\infty} f_n z^n, 
\end{align*}
we let $[z^n](f(z))=f_n$. Here noting
\begin{align*}
\frac{1}{w -\gamma} &= - \frac{1}{\gamma(1-w/\gamma)} = - \frac{1}{\gamma} \left(\frac{w}{\gamma}\right)^n,
\\
\frac{1}{w -\overline{\gamma}} &= - \frac{1}{\overline{\gamma}(1-w/\overline{\gamma})} = - \frac{1}{\overline{\gamma}} \left(\frac{w}{\overline{\gamma}}\right)^n, 
\end{align*}
we get
\begin{align*}
[w^n] \left( \frac{1}{w -\gamma} \right) &= - \gamma^{-(n+1)} = - e^{-i(n+1) \theta_0},
\\
[w^n] \left( \frac{1}{w -\overline{\gamma}} \right) &= - \overline{\gamma}^{-(n+1)} = - \gamma^{n+1} = - e^{i(n+1) \theta_0}.
\end{align*}
Therefore, we have
\begin{align*}
[w^n] \left( \frac{1-w^2}{(w - \gamma) (w - \overline{\gamma})} \right) 
&\sim 
[w^n] \left( \frac{1 - \gamma^2}{(\gamma - \overline{\gamma}) (w - \gamma)} \right) 
+ [w^n] \left( \frac{1 - \overline{\gamma}^2}{(\overline{\gamma} - \gamma) (w - \overline{\gamma})} \right).
\end{align*}
As for the above derivation, see pp.264-265 in Flajolet and Sedgewick \cite{Flajolet2009}, for example. By using Eq. \eqref{gotejobs1}, we obtain
\begin{align*}
[w^n] \left(\frac{1-w^2}{(w - \gamma) (w - \overline{\gamma})} \right) 
&\sim 
[w^n] \left( \frac{1- \gamma^2}{(\gamma - \overline{\gamma}) (w - \gamma)} + 
\frac{1 - \overline{\gamma}^2}{(\overline{\gamma} - \gamma) (w - \overline{\gamma})} \right)
\\
&= - \left\{ \frac{1 - \gamma^2}{\gamma - \overline{\gamma}} \> \gamma^{-(n+1)} + \frac{1 - \overline{\gamma}^2}{\overline{\gamma} - \gamma} \> \overline{\gamma}^{-(n+1)}\right\}
\\
&= - 2 \Re \left( \frac{1 - \gamma^2}{\gamma - \overline{\gamma}} \> \gamma^{-(n+1)} \right)
\\
&= - 2 \times \Re \left( \frac{\overline{\gamma} - \gamma}{\gamma - \overline{\gamma}} \> \gamma^{-n} \right)
\\
&= 2 \> \cos (n \theta_0).
\end{align*}
Thus, we have
\begin{align*}
[w^n] \left( A^{\alpha}_1 (w) \right) \sim \frac{1-\sqrt{2}S}{3-2 \sqrt{2}S} \> \cos (n \theta_0). 
\end{align*}
Similarly, we consider
\begin{align*}
A^{\beta}_1 (w) 
= \dfrac{-C \left\{1 + 2 (1- \sqrt{2}S)w +w^2 \right\}}{\sqrt{2}(3-2 \sqrt{2}S) \left\{ 1+ \dfrac{2(1-\sqrt{2}S)^2}{3-2 \sqrt{2}S}w +w^2\right\} }.
\end{align*}
Then, we get
\begin{align*}
&[w^n] \left( \frac{1 + 2 (1- \sqrt{2}S) w +w^2}{(w - \gamma) (w - \overline{\gamma})} \right) 
\\
&\sim [w^n] \left( \frac{1+ 2 (1- \sqrt{2}S) \gamma + \gamma^2}{(\gamma - \overline{\gamma}) (w - \gamma)} + 
\frac{1 + 2 (1- \sqrt{2}S) \overline{\gamma} + \overline{\gamma}^2}{(\overline{\gamma} - \gamma) (w - \overline{\gamma})} \right)
\\
&= - \left\{ \frac{1+ 2 (1- \sqrt{2}S) \gamma + \gamma^2}{\gamma - \overline{\gamma}} \> \gamma^{-(n+1)} + \frac{1 + 2 (1- \sqrt{2}S) \overline{\gamma} + \overline{\gamma}^2}{\overline{\gamma} - \gamma} \> \overline{\gamma}^{-(n+1)}\right\}
\\
&= -2 \Re \left( \frac{1+ 2 (1- \sqrt{2}S) \gamma + \gamma^2}{\gamma - \overline{\gamma}} \> \gamma^{-(n+1)} \right)
\\
&= - 2 \Re \left( \frac{\overline{\gamma} + 2(1- \sqrt{2}S) + \gamma}{\gamma - \overline{\gamma}} \> \gamma^{-n} \right)
\\
&= \frac{\sqrt{2}(1 - \sqrt{2}S)}{C} \> \sin (n \theta_0). 
\end{align*}
In the above derivation, we used
\begin{align*}
\sin (\theta_0) = \frac{2(\sqrt{2}-S)C}{3 - 2 \sqrt{2}S}.
\end{align*}
Therefore, we have
\begin{align*}
[w^n] \left(A^{\beta}_1 (w) \right) \sim - \frac{1 - \sqrt{2}S}{3 - 2 \sqrt{2}S} \> \sin (n \theta_0).
\end{align*}

Next we consider
\begin{align*}
A^{\alpha}_2 (w) 
=\frac{(\sqrt{2}-S)(1+w)\sqrt{1+w^2}}{\sqrt{2}(3-2 \sqrt{2}S) \left\{ 1+ \dfrac{2(1-\sqrt{2}S)^2}{3-2 \sqrt{2}S}w +w^2\right\} }.
\end{align*}
Then 
\begin{align*}
[w^n] \left( \frac{(1+w)\sqrt{1+w^2}}{(w - \gamma) (w - \overline{\gamma})} \right) 
&\sim - \left\{ \frac{(1+ \gamma) \sqrt{1+\gamma^2}}{\gamma - \overline{\gamma}} \> \gamma^{-(n+1)} 
\right.
\\
& \qquad \qquad \qquad \left.
- \frac{(1+ \overline{\gamma}) \sqrt{1+\overline{\gamma}^2}}{\overline{\gamma} - \gamma} \> \overline{\gamma}^{-(n+1)} \right\}
\\
&= - 2 \Re \left( \frac{(1+ \gamma) \sqrt{1+\gamma^2}}{\gamma - \overline{\gamma}} \> \gamma^{-(n+1)} \right)
\\
&=- 2\sqrt{-2 \cos \theta_0} \times \Re \left( \frac{(\overline{\gamma} +1) (-i) e^{i \theta_0/2}}{\gamma - \overline{\gamma}} \> \gamma^{-n} \right)
\\
&= \sqrt{-2 \cos \theta_0} \times \frac{\cos (n \theta_0)}{\sin(\theta_0/2)}.
\end{align*}
Here we should remark that
\begin{align*}
\sqrt{1+\gamma^2}
&= \sqrt{-2 \cos \theta_0} \left( \sin \left( \frac{\theta_0}{2} \right) - i \cos \left( \frac{\theta_0}{2} \right)\right) = \sqrt{-2 \cos \theta_0} (-i) e^{i \theta_0/2}, 
\\
\sqrt{1+\overline{\gamma}^2}
&= \overline{\sqrt{1+\gamma^2}}.
\end{align*}
Moreover, noting
\begin{align*}
\sqrt{-2 \cos \theta_0} = \frac{\sqrt{2}|1 - \sqrt{2}S|}{\sqrt{3-2 \sqrt{2}S}}, \quad \sin \left( \frac{\theta_0}{2} \right) = \frac{\sqrt{2}- S}{\sqrt{3-2 \sqrt{2}S}},
\end{align*}
we obtain
\begin{align*}
[w^n] \left( A^{\alpha}_2 (w) \right) \sim \frac{|1 - \sqrt{2}S|}{3-2 \sqrt{2}S} \cos (n \theta_0).
\end{align*}
In a similar fashion, we consider
\begin{align*}
A^{\beta}_2 (w) 
= \frac{C(1-w)\sqrt{1+w^2}}{\sqrt{2}(3-2 \sqrt{2}S) \left\{ 1+ \dfrac{2(1-\sqrt{2}S)^2}{3-2 \sqrt{2}S}w +w^2\right\} }.
\end{align*}
Then, we get
\begin{align*}
[w^n] \left( \frac{(1-w)\sqrt{1+w^2}}{(w - \gamma) (w - \overline{\gamma})} \right) 
&\sim - \left\{ \frac{(1- \gamma) \sqrt{1+\gamma^2}}{\gamma - \overline{\gamma}} \> \gamma^{-(n+1)} 
\right.
\\
& \qquad \qquad \qquad \left.
+ \frac{(1- \overline{\gamma}) \sqrt{1+\overline{\gamma}^2}}{\overline{\gamma} - \gamma} \> \overline{\gamma}^{-(n+1)} \right\}
\\
&= - 2 \Re \left( \frac{(1- \gamma) \sqrt{1+\gamma^2}}{\gamma - \overline{\gamma}} \> \gamma^{-(n+1)} \right)
\\
&= - 2 \sqrt{-2 \cos \theta_0} \> \Re \left( \frac{(\overline{\gamma} - 1) (-i) e^{i \theta_0/2}}{\gamma - \overline{\gamma}} \> \gamma^{-n} \right)
\\
&= - \sqrt{-2 \cos \theta_0} \times \frac{\sin (n \theta_0)}{\cos (\theta_0/2)} 
\\
&= - \frac{\sqrt{2}|1- \sqrt{2}S|}{C} \sin (n \theta_0)
\end{align*}
In this derivation, we used 
\begin{align*}
\sqrt{-2 \cos \theta_0} = \frac{\sqrt{2}|1 - \sqrt{2}S|}{\sqrt{3-2 \sqrt{2}S}}, \quad \cos \left( \frac{\theta_0}{2} \right) = \frac{C}{\sqrt{3-2 \sqrt{2}S}}.
\end{align*}
Thus, we have
\begin{align*}
[w^n] \left( A^{\beta}_2 (w) \right) \sim - \frac{|1- \sqrt{2}S|}{3-2 \sqrt{2}S} \sin (n \theta_0).
\end{align*}
Therefore, we obtain
\begin{align*}
\Psi_{2n} ^{L} (0) 
&= [w^n] \left\{  (A^{\alpha} _1 (w) + A^{\alpha} _2 (w)) \> \alpha + (A^{\beta} _1 (w) + A^{\beta} _2 (w)) \> \beta \right\}
\\
&= [w^n] \{ A^{\alpha} _1 (w) + A^{\alpha} _2 (w) \} \> \alpha +  [w^n] \{ A^{\beta} _1 (w) + A^{\beta} _2 (w)) \} \> \beta 
\\
&\sim \frac{2(1- \sqrt{2}S)}{3-2 \sqrt{2}S} \> \left\{ \cos (n \theta_0) \> \alpha - \sin (n \theta_0) \> \beta \right\} \times I_{[0, \pi/4)}(\xi).
\end{align*}
So the proof of Theorem \ref{gotennba} is complete.

\section{Space-time generating function method \label{bokannsuu}}

In general, both the Fourier analysis and stationary phase method are useful for investigating space-homogeneous QWs, however, not for space-inhomogeneous models. On the other hand, the space-time generating function method introduced by \cite{KLS2013} is applicable to some space-inhomogeneous QWs. The result given here can be obtained as a corollary of the result in \cite{KLS2013}. However, for the convenience for readers, we will explain the details on this derivation in this paper.

First, the quantum coin $U_x$ at position $x$ is given by 
\begin{align*}
U_x= 
\begin{bmatrix} 
a_x & b_x \\ c_x & d_x 
\end{bmatrix}. 
\end{align*}
Here we suppose that $a_x b_x c_x d_x \not= 0$ for simplicity. Let $\Delta_x = \det U_x$. 

Let $F^{(+)}(x,n)$ denote the sum of all path, which the quantum walker starting from position $x$ moves in the region $\{ y \in \mathbb{Z}: y \ge x \}$ and reaches position $x$ at time $n$ for the first time. For example, 
\begin{align*}
F^{(+)}(x,2) = P_{x+1} Q_{x}, \quad F^{(+)}(x,4) = P_{x+1} P_{x+2} Q_{x+1} Q_{x}. 
\end{align*}
Indeed, each path has the form $P_{x+1} \cdots Q_{x}$, so $F^{(+)}(x,n)$ is expressed that there exists $f^{(+)}(x,n) (\in \CM)$ such that
\begin{align}
F^{(+)}(x,n) = f^{(+)}(x,n) R_x,
\label{timosi}
\end{align}
where
\begin{align*}
R_x = \begin{bmatrix} c_x & d_x \\ 0 & 0 \end{bmatrix}. 
\end{align*}
In fact, noting
\begin{align*}
F^{(+)}(x,2) &= P_{x+1} Q_{x} = b_{x+1} R_x, 
\\
F^{(+)}(x,4) &= P_{x+1} P_{x+2} Q_{x+1} Q_{x} = a_{x+1} d_{x+1} b_{x+2} R_x,
\end{align*}
we have $f^{(+)}(x,2) = b_{x+1}, \> f^{(+)}(x,4) = a_{x+1} d_{x+1} b_{x+2}$. Here we introduce the generating function of $F^{(+)}(x,n)$ with respect to time $n$ as follows.
\begin{align*}
\widetilde{F}^{(+)}_x(z) = \sum_{n=2}^{\infty} F^{(+)}(x,n) z^n.
\end{align*}
Moreover, we put
\begin{align*}
\fp_x(z) = \sum_{n=2}^{\infty} f^{(+)}(x,n) z^n.
\end{align*}
We should remark that $\widetilde{F}^{(+)}_x(0) =O_2, \> \fp_x(0) = 0$, where $O_n$ is the $n \times n$ zero matrix. Thus, Eq.\eqref{timosi} gives
\begin{align*}
\widetilde{F}^{(+)}_x(z) = \sum_{n=2}^{\infty} f^{(+)}(x,n) R_x z^n = \fp_x(z) R_x.
\end{align*}
Indeed, we have
\begin{align}
\widetilde{F}^{(+)}_x(z) = \fp_x(z) \begin{bmatrix} c_x & d_x \\ 0 & 0 \end{bmatrix}. 
\label{riari}
\end{align}

Similarly, $F^{(-)}(x,n)$ denotes the sum of all path, which the quantum walker starting from position $x$ moves in the region $\{ y \in \mathbb{Z}: y \le x \}$ and reaches position $x$ at time $n$ for the first time. For example, 
\begin{align*}
F^{(-)}(x,2) = Q_{x-1} P_{x}, \quad F^{(-)}(x,4) = Q_{x-1} Q_{x-2} P_{x-1} P_{x}. 
\end{align*}
In fact, each path has the form $P_{x} \cdots Q_{x-1}$, so $F^{(-)}(x,n)$ is expressed that there exists $f^{(-)}(x,n) (\in \CM)$ such that 
\begin{align}
F^{(-)}(x,n) = f^{(-)}(x,n) S_x,
\label{timosi2}
\end{align}
where
\begin{align*}
S_x = \begin{bmatrix} 0 & 0 \\ a_x & b_x \end{bmatrix}. 
\end{align*}
For instance, we have
\begin{align*}
F^{(-)}(x,2) &=  Q_{x-1} P_{x} = c_{x-1} S_x, 
\\
F^{(-)}(x,4) &= Q_{x-1} Q_{x-2} P_{x-1} P_{x} = a_{x-1} d_{x-1} c_{x-2} S_x.
\end{align*}
So, we get $f^{(-)}(x,2) = c_{x-1}, \> f^{(-)}(x,4) = a_{x-1} d_{x-1} c_{x-2}$. Here we introduce a generating function of $F^{(-)}(x,n)$ with respect to time $n$:
\begin{align*}
\widetilde{F}^{(-)}_x(z) = \sum_{n=2}^{\infty} F^{(-)}(x,n) z^n.
\end{align*}
Furthermore, 
\begin{align*}
\fm_x(z) = \sum_{n=2}^{\infty} f^{(-)}(x,n) z^n.
\end{align*}
Remark that $\widetilde{F}^{(-)}_x(0) = O_2, \> \fm_x(0) = 0$. Thus, Eq. \eqref{timosi2}, we have
\begin{align*}
\widetilde{F}^{(-)}_x(z) = \sum_{n=2}^{\infty} f^{(-)}(x,n) S_x z^n = \fm_x(z) S_x.
\end{align*}
In fact, we get
\begin{align}
\widetilde{F}^{(-)}_x(z) = \fm_x(z) \begin{bmatrix} 0 & 0 \\ a_x & b_x \end{bmatrix}. 
\label{riari2}
\end{align}

Let $\Xi^{(+)}(x,n)$ denote the sum of all path, which the quantum walker starting from position $x$ moves in the region $\{ y \in \mathbb{Z}: y \ge x \}$ and reaches position $x$ at time $n$. For example, 
\begin{align*}
\Xi^{(+)}(x,2) &= P_{x+1} Q_{x}, 
\\
\Xi^{(+)}(x,4) &= P_{x+1} P_{x+2} Q_{x+1} Q_{x} + P_{x+1} Q_{x} P_{x+1} Q_{x}. 
\end{align*}
Here, we introduce a generating function of $\Xi^{(+)}(x,n)$ with respect to time $n$: 
\begin{align*}
\widetilde{\Xi}^{(+)}_x(z) = \sum_{n=0}^{\infty} \Xi^{(+)}(x,n) z^n.
\end{align*}
From now on, we consider a relation between $\widetilde{\Xi}^{(+)}_x(z)$ and $\widetilde{F}^{(+)}_x(z)$. Each definition implies 
\begin{align*}
\Xi^{(+)}(x,2) &= F^{(+)}(x,2), 
\\
\Xi^{(+)}(x,4) &= F^{(+)}(x,4) + F^{(+)}(x,2)^2,
\\
\Xi^{(+)}(x,6) &= F^{(+)}(x,6) + F^{(+)}(x,4) F^{(+)}(x,2) 
\\
& \qquad \qquad + F^{(+)}(x,2) F^{(+)}(x,4) + F^{(+)}(x,2)^3.
\end{align*}
Thus, we have
\begin{align*}
\widetilde{\Xi}^{(+)}_x(z)= I + \widetilde{F}^{(+)}_x(z) + \left(\widetilde{F}^{(+)}_x(z)\right)^2 + \cdots = I+ \widetilde{F}^{(+)}_x(z) \widetilde{\Xi}^{(+)}_x(z).
\end{align*}
Here, $I=I_2$ is the $2 \times 2$ identity matrix. Therefore, we get
\begin{align*}
\widetilde{\Xi}^{(+)}_x(z) = (I - \widetilde{F}^{(+)}_x(z))^{-1}.
\end{align*}
From Eq. \eqref{riari}, we have
\begin{align}
\label{pero1}
\widetilde{\Xi}^{(+)}_x(z) = \frac{1}{1-c_x \fp_x(z)} 
\begin{bmatrix} 1 & d_x \fp_x(z) \\ 0 & 1-c_x \fp_x(z) \end{bmatrix}. 
\end{align}	
Moreover, we obtain 
\begin{align}
\label{pero2}
\widetilde{F}^{(+)}_x(z)=zP_{x+1} \; \widetilde{\Xi}^{(+)}_{x+1}(z)\; zQ_x.
\end{align}
Combining Eq. \eqref{pero2} with Eqs. \eqref{riari} and \eqref{pero1} gives
\begin{align*}
\fp_x(z) = \frac{z^2 (\Delta_{x+1}\fp_{x+1} (z) + b_{x+1})}{1-c_{x+1} \fp_{x+1}(z)}. 
\end{align*}
By this equation, we have the following continued fraction expansion on $\{\fp_y (z) : y=x, x+1, x+2, \ldots \}$:
\begin{align} 
\fp_x (z) = -\frac{z^2\Delta_{x+1}}{c_{x+1}}\left( 1-\frac{|a_{x+1}|^2}{1-c_{x+1}\fp_{x+1}(z)} \right).
\label{pero+}
\end{align}

In a similar fashion, let $\Xi^{(-)}(x,n)$ denote the sum of all path, which the quantum walker starting from position $x$ moves in the region $\{ y \in \mathbb{Z}: y \le x \}$ and reaches position $x$ at time $n$. For example, 
\begin{align*}
\Xi^{(-)}(x,2) &= Q_{x-1} P_{x}, 
\\
\Xi^{(-)}(x,4) &= Q_{x-1} Q_{x-2} P_{x-1} P_{x} + Q_{x-1} P_{x} Q_{x-1} P_{x}.
\end{align*}
Here, we introduce a generating function of $\Xi^{(-)}(x,n)$ with respect to time $n$ as follows.
\begin{align*}
\widetilde{\Xi}^{(-)}_x(z) = \sum_{n=0}^{\infty} \Xi^{(-)}(x,n) z^n.
\end{align*}
As in the case of $\Xi^{(+)}(x,n)$, we have 
\begin{align*}
\widetilde{\Xi}^{(-)}_x(z)= I + \widetilde{F}^{(-)}_x(z) + \left(\widetilde{F}^{(-)}_x(z)\right)^2 + \cdots = I+ \widetilde{F}^{(-)}_x(z) \widetilde{\Xi}^{(-)}_x(z).
\end{align*}
Thus, we see 
\begin{align*}
\widetilde{\Xi}^{(-)}_x(z) = (I - \widetilde{F}^{(-)}_x(z))^{-1}.
\end{align*}
From Eq. \eqref{riari2}, we get
\begin{align}
\label{pero33}
\widetilde{\Xi}^{(-)}_x(z)
=\frac{1}{1-b_x \fm_x(z)}
\begin{bmatrix} 1-b_x\fm_x(z) & 0 \\ a_x \fm_x(z) & 1 \end{bmatrix}.
\end{align}	
Moreover, we obtain
\begin{align}
\label{pero4}
\widetilde{F}^{(-)}_x(z)=zQ_{x-1}\; \widetilde{\Xi}^{(-)}_{x-1}(z)\; zP_x.
\end{align}
Combining Eq. \eqref{pero4} with Eqs. \eqref{riari2} and \eqref{pero33} gives
 the following continued fraction expansion on $\{\fm_y (z) : y=x, x-1, x-2, \ldots \}$:
\begin{align} 
\fm_x(z) = -\frac{z^2\Delta_{x-1}}{b_{x-1}}
\left( 1-\frac{|d_{x-1}|^2}{1-b_{x-1}\fm_{x-1}(z)} \right).
\label{pero-}
\end{align}

Let $\Xi(x,n)$ denote the sum of all path, which the quantum walker starting from the origin reaches position $x$ at time $n$. We introduce a generating function of $\Xi(x,n)$ with respect to time $n$ as follows.
\begin{align*}
\widetilde{\Xi}_x (z) = \sum_{n=0}^{\infty} \Xi(x,n) z^n. 
\end{align*}

First, we consider $x=0$ case. A relation between $\widetilde{\Xi}_0(z)$ and $\widetilde{F}_0^{(\pm)}(z)$ implies
\begin{align*}
\widetilde{\Xi}_0(z) 
&=I+\left(\widetilde{F}_0^{(+)}(z)+\widetilde{F}_0^{(-)}(z)\right)+\left(\widetilde{F}_0^{(+)}(z)+\widetilde{F}_0^{(-)}(z)\right)^2+\cdots
\\
&=I+\left(\widetilde{F}_0^{(+)}(z)+\widetilde{F}_0^{(-)}(z)\right)\widetilde{\Xi}_0(z). 
\end{align*}
By Eqs. \eqref{riari} and \eqref{riari2}, we have
\begin{align*}
\widetilde{F}_0^{(+)}(z)+\widetilde{F}_0^{(-)}(z) =  \begin{bmatrix} c_0 \fp_0(z) & d_0 \fp_0(z) \\ a_0 \fm_0(z) & b_0 \fm_0(z) \end{bmatrix}. 
\end{align*}
Thus, we see
\begin{align}
\label{pero0}
\widetilde{\Xi}_0(z) 
= \frac{1}{\gamma (z)}
\begin{bmatrix} 
1-b_0\fm_0(z) & d_0\fp_0(z) \\ a_0\fm_0(z) & 1-c_0\fp_0(z)
\end{bmatrix},
\end{align}
where
\begin{align*}
\gamma (z) = 1-c_0\fp_0(z)-b_0\fm_0(z)-\Delta_0\fp_0(z)\fm_0(z).
\end{align*}

Next, we consider $x \ge 1$ case. Then we see
\begin{align}
\label{pero3}
\widetilde{\Xi}_x(z)=\widetilde{\Xi}_x^{(+)}(z)\;zQ_{x-1}\;\widetilde{\Xi}_{x-1}(z).   
\end{align}
For $0 \le y \le x$, we put
\begin{align*}
|u^{(+)}_y \rangle = 
\begin{bmatrix} 
\lp_y(z)\fp_y(z) \\ z
\end{bmatrix},
\quad
|v^{(+)}_y \rangle =
\begin{bmatrix} 
\overline{c}_y \\ \overline{d}_y
\end{bmatrix},
\end{align*}
where
\begin{align}
\lp_y(z) = \frac{zd_y}{1-c_y\fp_y(z)}.
\label{ponta1}
\end{align}
Then we have
\begin{align}
\langle v^{(+)}_{y} ,u^{(+)}_{y} \rangle = \lp_y(z).
\label{march1}
\end{align}
From Eq. \eqref{pero1}, we get
\begin{align}
\widetilde{\Xi}_x^{(+)}(z) \; zQ_{x-1} = |u^{(+)}_x\rangle \langle v^{(+)}_{x-1}|.
\label{march2}
\end{align}
By using Eqs. \eqref{pero3}�C\eqref{march1}�Cand \eqref{march2}, we see that $\widetilde{\Xi}_x(z)$ can be expressed as  
\begin{align*}
\widetilde{\Xi}_x(z) 
&= |u^{(+)}_x\rangle \langle v^{(+)}_{x-1}|\widetilde{\Xi}_{x-1}(z)
\\
&= |u^{(+)}_x \rangle \langle v^{(+)}_{x-1} ,u^{(+)}_{x-1} \rangle \cdots \langle v^{(+)}_1,u^{(+)}_1 \rangle \langle v^{(+)}_0| \widetilde{\Xi}_0(z) 
\\
&= \langle v^{(+)}_{x-1},u^{(+)}_{x-1} \rangle \cdots \langle v^{(+)}_1,u^{(+)}_1 \rangle |u^{(+)}_x \rangle \langle v^{(+)}_0| \widetilde{\Xi}_0(z) 
\\
&= \lp_{x-1}(z)\cdots\lp_{1}(z)
\begin{bmatrix} \lp_{x}(z)\fp_x(z) \\ z \end{bmatrix}
\begin{bmatrix} c_0 & d_0 \end{bmatrix} \widetilde{\Xi}_0(z).
\end{align*}
Here, for $x=1$ case, we have
\begin{align*}
\widetilde{\Xi}_1 (z) = 
\begin{bmatrix} \lp_{1}(z)\fp_1(z) \\ z \end{bmatrix}
\begin{bmatrix} c_0 & d_0 \end{bmatrix}\widetilde{\Xi}_0(z).
\end{align*}

Similarly, for $x \le -1$ case, we have the corresponding results in the following way. First, we should remark
\begin{align}
\label{pero3m}
\widetilde{\Xi}_x(z)=\widetilde{\Xi}_x^{(-)}(z)\;zP_{x+1}\;\widetilde{\Xi}_{x+1}(z).   
\end{align}
Furthermore, for $x \le y \le 0$, we put
\begin{align*}
|u^{(-)}_y \rangle = 
\begin{bmatrix} 
z \\ \lm_y(z) \fm_y(z)
\end{bmatrix},
\quad
|v^{(-)}_y \rangle =
\begin{bmatrix} 
\overline{a}_y \\ \overline{b}_y
\end{bmatrix},
\end{align*}
where
\begin{align}
\lm_y(z) = \frac{za_y}{1-b_y\fm_y(z)}.
\label{ponta2}
\end{align}
Then we have
\begin{align}
\langle v^{(-)}_{y} ,u^{(-)}_{y} \rangle = \lm_y(z).
\label{march1m}
\end{align}
From Eq. \eqref{pero33}, we get
\begin{align}
\widetilde{\Xi}_x^{(-)}(z)\;zP_{x+1} = |u^{(-)}_x \rangle \langle v^{(-)}_{x+1}|.
\label{march2m}
\end{align}
By using Eqs. \eqref{pero3m}, \eqref{march1m}, and \eqref{march2m}, we see that $\widetilde{\Xi}_x(z)$ can be expressed as 
\begin{align*}
\widetilde{\Xi}_x(z) 
&= |u^{(-)}_x\rangle \langle v^{(-)}_{x+1}|\widetilde{\Xi}_{x+1}(z)
\\
&= |u^{(-)}_x \rangle \langle v^{(-)}_{x+1} ,u^{(-)}_{x+1} \rangle \cdots \langle v^{(-)}_{-1},u^{(-)}_{-1} \rangle \langle v^{(+)}_0| \widetilde{\Xi}_0(z) 
\\
&= \langle v^{(-)}_{x+1},u^{(-)}_{x+1} \rangle \cdots \langle v^{(-)}_{-1},u^{(-)}_{-1} \rangle |u^{(-)}_x \rangle \langle v^{(-)}_0| \widetilde{\Xi}_0(z) 
\\
&= \lm_{x+1}(z)\cdots\lp_{-1}(z)
\begin{bmatrix} z \\ \lm_{x}(z)\fm_x(z) \end{bmatrix}
\begin{bmatrix} a_0 & b_0 \end{bmatrix}\widetilde{\Xi}_0(z).
\end{align*}
Here, for $x=-1$ case, we have
\begin{align*}
\widetilde{\Xi}_{-1} (z) = 
\begin{bmatrix} z \\ \lm_{-1}(z)\fp_{-1}(z) \end{bmatrix}
\begin{bmatrix} a_0 & b_0 \end{bmatrix}\widetilde{\Xi}_0(z).
\end{align*}

Therefore, we obtain
\begin{pro} 
\label{acco} Put $\Delta_x = \det(U_x)$. Then 
\begin{enumerate} 
        \item if $x=0$, 
	\begin{equation*}\label{pero0}
		\widetilde{\Xi}_0(z) 
		= \frac{1}{\gamma (z)}
        		\begin{bmatrix} 1-b_0\fm_0(z) & d_0\fp_0(z) \\ a_0\fm_0(z) & 1-c_0\fp_0(z)\end{bmatrix},
	\end{equation*}
        \item if $|x|\geq 1$, 
        \begin{equation*} 
		\widetilde{\Xi}_x(z)
		= \begin{cases}
                	\lp_{x-1}(z)\cdots\lp_{1}(z)
                 		\begin{bmatrix} \lp_{x}(z)\fp_x(z) \\ z\end{bmatrix}
                        	\begin{bmatrix} c_0 & d_0 \end{bmatrix}\widetilde{\Xi}_0(z) & \text{$(x\geq 1)$, } \\
                        	\\
                 	\lm_{x+1}(z)\cdots\lm_{-1}(z)
                 		\begin{bmatrix} z \\ \lm_{x}(z)\fm_x(z) \end{bmatrix}
                        	\begin{bmatrix} a_0 & b_0 \end{bmatrix}\widetilde{\Xi}_0(z) & \text{$(x\leq -1)$,}       
        	  \end{cases}    
         \end{equation*}
\end{enumerate} 
where
\begin{align*}
\gamma (z) 
&= 1-c_0\fp_0(z)-b_0\fm_0(z)-\Delta_0\fp_0(z)\fm_0(z),
\\
\lp_x(z) &=zd_x/(1-c_x\fp_x(z)), \quad \lm_x(z)=za_x/(1-b_x\fm_x(z)).
\end{align*}
Here, $\widetilde{f}^{(\pm)}_x(z)$ has the following continued fraction expansion:
\begin{align*} 
	\fp_x(z) &= -\frac{z^2\Delta_{x+1}}{c_{x+1}}\left( 1-\frac{|a_{x+1}|^2}{1-c_{x+1}\fp_{x+1}(z)} \right), \\
        \fm_x(z) &= -\frac{z^2\Delta_{x-1}}{b_{x-1}}\left( 1-\frac{|d_{x-1}|^2}{1-b_{x-1}\fm_{x-1}(z)} \right). 
\end{align*}
\end{pro}
\section{Time-averaged limit measure \label{sectalm}}

From the definition of our model and continued fraction expansions of $\widetilde{f}_x ^{(\pm)} (z)$, we see that $\fp_0(z) = \fm_0(z) = \fp_x(z) = \fm_x(z)$, so we put $\widetilde{f}_0 (z) = \fp_0(z) $. Thus, Eq. \eqref{pero+} gives 
\begin{align*}
\widetilde{f}_0 (z) 
= - \frac{z^2 \Delta_1}{c_1} \left(1 - \frac{|a_1|^2}{1-c_1 \widetilde{f}_0 (z) } \right)
= \sqrt{2} z^2 \left(1 - \frac{1}{2 - \sqrt{2} \widetilde{f}_0 (z) } \right).
\end{align*}
Therefore, $\widetilde{f}_0 (z)$ is a solution of  
\begin{align*}
x^2 - \sqrt{2} (z^2+1) x + z^2 =0.
\end{align*}
By definition, we have $\widetilde{f}_0 (0) = 0$. Thus, we have
\begin{align*}
\widetilde{f}_0 (z) = \frac{z^2+1 - \sqrt{z^4+1}}{\sqrt{2}}.
\end{align*}
Put $z=e^{i \theta}$. So we get
\begin{align}
\widetilde{f}_0 (e^{i \theta}) = e^{i \theta} \left( \sqrt{2} \cos \theta + i {\rm sgn} (\sin \theta) \sqrt{1 - 2 \cos^2 \theta} \right),
\label{1Dhardnut1Gen}
\end{align}
where $\theta \in [-3 \pi/4, - \pi/4) \cup  [\pi/4, 3 \pi/4)$. Furthermore, if we let 
\begin{align*}
\widetilde{f}_0 (e^{i \theta}) = e^{i (\theta + \widetilde{\phi} (\theta))},
\end{align*}
Eq. \eqref{1Dhardnut1Gen} gives
\begin{align}
\cos ( \widetilde{\phi} (\theta)) 
&= \sqrt{2} \cos \theta, 
\label{1Dhardnut2Gen}
\\
\sin ( \widetilde{\phi} (\theta)) 
&= {\rm sgn} (\sin \theta) \sqrt{1 - 2 \cos^2 \theta}.
\label{1Dhardnut3}
\end{align}
On the other hand, $\widetilde{f}_0 (z) = \fp_0(z)=\fm_0(z)$ implies
\begin{align*}
\gamma (z) 
&= 1-c_0\fp_0(z)-b_0\fm_0(z)-\Delta_0\fp_0(z)\fm_0(z)
\\
&= 1 - 2 S \widetilde{f}_0 (z) + \widetilde{f}_0 (z)^2.
\end{align*}
Thus, we see that
\begin{align*}
\gamma (z) = 1 - 2 S \widetilde{f}_0 (z) + \left( \widetilde{f}_0 (z) \right)^2 = 0
\end{align*}
is equivalent to
\begin{align}
\widetilde{f}_0 (z) = e^{i(\theta + \widetilde{\phi} (\theta))} = S \pm C i.
\label{1Dtaihu27Gen}
\end{align}
So we consider the following case:
\begin{align}
e^{i(\theta + \widetilde{\phi} (\theta))} = S + Ci.
\label{1Dtaihu28}
\end{align}
Thus, we have
\begin{align*}
\cos (\theta + \widetilde{\phi} (\theta))= S. 
\end{align*}
So 
\begin{align}
\sqrt{2} \cos^2 \theta - \sin \theta \> \sin (\widetilde{\phi} (\theta)) = S. 
\label{kamakura1Gen}
\end{align}
Here, we used $\cos (\widetilde{\phi} (\theta))= \sqrt{2} \cos \theta$. On the other hand,
\begin{align*}
\sin (\theta + \widetilde{\phi} (\theta))= C.
\end{align*}
Thus, 
\begin{align}
\sqrt{2} \sin \theta \> \cos \theta + \cos \theta \> \sin (\widetilde{\phi} (\theta)) = C. 
\label{kamakura2Gen}
\end{align}
By using Eqs. \eqref{kamakura1Gen} and \eqref{kamakura2Gen}, we have 
\begin{align}
\sqrt{2} \cos \theta = S \cos \theta + C \sin \theta.
\label{minzoku2Gen}
\end{align}
Thus, from this equation, we have
\begin{align}
\cos \theta = \frac{C}{\sqrt{2}-S} \> \sin \theta.
\end{align}
Therefore, noting
\begin{align}
\sin \theta = \pm \frac{\sqrt{2}-S}{\sqrt{3 - 2 \sqrt{2}S}},
\end{align}
we get $\cos \theta = \pm C/\sqrt{3 - 2 \sqrt{2}S}$. Then, we see that two solutions of $\gamma (z)=0$ with $|z|=1$ are as follows:
\begin{align}
\left( \cos \theta^{(1)}, \sin \theta^{(1)} \right) 
&= \left(\frac{C}{\sqrt{3 - 2 \sqrt{2}S}}, \frac{\sqrt{2}-S}{\sqrt{3 - 2 \sqrt{2}S}} \right),
\label{1Ddoseki1Gen}
\\
\left( \cos \theta^{(2)}, \sin \theta^{(2)} \right) 
&= \left(-\frac{C}{\sqrt{3 - 2 \sqrt{2}S}}, -\frac{\sqrt{2}-S}{\sqrt{3 - 2 \sqrt{2}S}} \right),
\label{1Ddoseki2Gen}
\end{align}
where $\theta^{(1)}, \theta^{(2)} \in [-3 \pi/4, - \pi/4) \cup [\pi/4, 3\pi/4)$.

Next, we consider the following case:
\begin{align*}
e^{i(\theta + \widetilde{\phi} (\theta))} = S-Ci. 
\end{align*}
In a similar fashion, noting $\sqrt{2} \cos \theta = S \cos \theta - C \sin \theta$, we have
\begin{align}
\left( \cos \theta^{(3)}, \sin \theta^{(3)} \right) 
&= \left(\frac{C}{\sqrt{3 - 2 \sqrt{2}S}}, - \frac{\sqrt{2}-S}{\sqrt{3 - 2 \sqrt{2}S}} \right),
\label{1Ddoseki3Gen}
\\
\left( \cos \theta^{(4)}, \sin \theta^{(4)} \right) 
&= \left(-\frac{C}{\sqrt{3 - 2 \sqrt{2}S}}, \frac{\sqrt{2}-S}{\sqrt{3 - 2 \sqrt{2}S}} \right),
\label{1Ddoseki4Gen}
\end{align}
where $\theta^{(3)}, \theta^{(4)} \in [-3 \pi/4, - \pi/4) \cup [\pi/4, 3\pi/4)$.

From now on, we compute the residue. We should remark that 
\begin{align*}
\widetilde{\Xi}_0(z) 
&= \frac{1}{1 - b_0 \widetilde{f}_0(z) -c_0 \widetilde{f}_0(z) -\Delta_0 \widetilde{f}_0(z)^2}
\begin{bmatrix} 
1-b_0 \widetilde{f}_0(z) & d_0 \widetilde{f}_0(z) \\ 
a_0 \widetilde{f}_0(z) & 1-c_0 \widetilde{f}_0(z)
\end{bmatrix}
\\
&= \frac{1}{1 - 2 S \widetilde{f}_0 + \widetilde{f}_0 (z)^2}
\begin{bmatrix} 
1 - S \widetilde{f}_0(z) & - C \widetilde{f}_0(z) \\ 
C \widetilde{f}_0(z) & 1 - S \widetilde{f}_0(z)
\end{bmatrix}.
\end{align*}
Thus, we have
\begin{align}
\widetilde{\Xi}_0(z) \varphi
=
\widetilde{\Xi}_0(z)
\begin{bmatrix} 
\alpha \\
\beta
\end{bmatrix} 
=
\frac{1}{\gamma (z)}
\begin{bmatrix} 
(1 - S \widetilde{f}_0(z)) \alpha - C \widetilde{f}_0(z) \beta
\\ 
C \widetilde{f}_0(z) \alpha + (1 - S \widetilde{f}_0(z)) \beta
\end{bmatrix},
\label{1DtaihendaGen}
\end{align}
where
\begin{align*}
\gamma (z) = 1 - 2 S \widetilde{f}_0 (z) + \left( \widetilde{f}_0 (z) \right)^2.
\end{align*}
Putting $z=e^{i \theta}$, we get
\begin{align*}
\frac{\partial \gamma (z)}{\partial z} 
= - i e^{- i \theta} \frac{\partial \gamma (e^{i \theta})}{\partial \theta}.
\end{align*}
By using 
\begin{align*}
\gamma (e^{i \theta}) = 1 - 2 S  e^{i(\theta + \widetilde{\phi} (\theta))} + e^{2i(\theta + \widetilde{\phi} (\theta))},
\end{align*}
we have
\begin{align}
\frac{\partial \gamma (e^{i \theta})}{\partial \theta}
&= 2 i \left( 1 + \frac{\partial \widetilde{\phi}(e^{i \theta})}{\partial \theta} \right)  e^{i(\theta + \widetilde{\phi} (\theta))} \left( -S + e^{i(\theta + \widetilde{\phi} (\theta))} \right).
\label{1Dtezukaosamu1Gen}
\end{align}

We should note that
\begin{align*}
&\left\| {\rm Res} \left( \widetilde{\Xi}_0 (z) \varphi ; z=e^{i \theta} \right) \right\|^2
\\
& \qquad \qquad =
\left| {\rm Res} \left( \frac{(1 - S \widetilde{f}_0(z)) \alpha - C \widetilde{f}_0(z) \beta}{\gamma (z)} ; z=e^{i \theta} \right) \right|^2
\\
& \qquad \qquad + 
\left| {\rm Res} \left( \frac{C \widetilde{f}_0(z) \alpha + (1 - S \widetilde{f}_0(z)) \beta}{\gamma (z)} ; z=e^{i \theta} \right) \right|^2.
\end{align*}
First, we consider $z=e^{i \theta^{(1)}}$ case. Then we see 
\begin{align}
&\left| {\rm Res} \left( \frac{(1 - S \widetilde{f}_0(z)) \alpha - C \widetilde{f}_0(z) \beta}{\gamma (z)} ; z=e^{i \theta^{(1)}} \right) \right|^2
\nonumber
\\
& \qquad \qquad =
\frac{\left| (1 - S \widetilde{f}_0 (e^{i \theta^{(1)}})) \alpha - C \widetilde{f}_0(e^{i \theta^{(1)}}) \beta \right|^2}{\left|\dfrac{\partial \gamma (e^{i \theta^{(1)}})}{\partial \theta^{(1)}} \right|^2},
\nonumber
\\
&\left| {\rm Res} \left( \frac{C \widetilde{f}_0(z) \alpha + (1 - S \widetilde{f}_0(z)) \beta}{\gamma (z)} ; z=e^{i \theta^{(1)}} \right) \right|^2
\nonumber
\\
& \qquad \qquad =
\frac{\left| C \widetilde{f}_0(z) \alpha + (1 - S \widetilde{f}_0(z)) \beta \right|^2}{\left|\dfrac{\partial \gamma (e^{i \theta^{(1)}})}{\partial \theta^{(1)}} \right|^2},
\label{1Dbudda1Gen}
\end{align}
where
\begin{align*}
\dfrac{\partial \gamma (e^{i \theta^{(1)}})}{\partial \theta^{(1)}}= \left. \dfrac{\partial \gamma (e^{i \theta})}{\partial \theta} \right|_{\theta=\theta^{(1)}}.
\end{align*}
Here we compute 
\begin{align*}
\frac{\partial \widetilde{\phi}(e^{i \theta^{(1)}})}{\partial \theta^{(1)}}
&= \left. \frac{\partial \widetilde{\phi}(e^{i \theta^{(1)}})}{\partial \theta} \right|_{\theta = \theta^{(1)}} = \frac{\sqrt{2} \sin \theta^{(1)}}{\sin \widetilde{\phi}(\theta^{(1)})} = \sqrt{2} \times \frac{\sqrt{2}-S}{|1 - \sqrt{2}S|}.
\end{align*}
The second equality comes from Eq. \eqref{1Dhardnut2Gen}. Thus, we have 
\begin{align}
\left| \frac{\partial \gamma (e^{i \theta^{(1)}})}{\partial \theta^{(1)}} \right|^2
= 
4  C^2 \left| 1 + \frac{\partial \widetilde{\phi}(e^{i \theta^{(1)}})}{\partial \theta^{(1)}} \right|^2
= 4 C^2 \frac{(3 - 2 \sqrt{2}S)^2}{(1 - \sqrt{2}S)^2}.
\label{1Dbudda2Gen}
\end{align}
From Eqs. \eqref{1Dbudda1Gen} and \eqref{1Dbudda2Gen}, we get
\begin{align*}
&\left\| {\rm Res} \left( \widetilde{\Xi}_0 (z) \varphi ; z=e^{i \theta^{(1)}} \right) \right\|^2
\\
& \quad = \frac{\left|(1 - S \widetilde{f}_0 (e^{i \theta^{(1)}})) \alpha - C \widetilde{f}_0(e^{i \theta^{(1)}}) \beta \right|^2 + \left|C \widetilde{f}_0(e^{i \theta^{(1)}}) \alpha + (1 - S \widetilde{f}_0 (e^{i \theta^{(1)}})) \beta \right|^2}{\left| \dfrac{\partial \gamma (e^{i \theta^{(1)}})}{\partial \theta^{(1)}} \right|^2}
\\
& \quad = \frac{(1 - \sqrt{2}S)^2}{2(3 - 2 \sqrt{2}S)^2} \times |\alpha - i \beta|^2.
\end{align*}
In this derivation, we used $\widetilde{f}_0(e^{i \theta^{(1)}})=S+Ci$. Similarly, we have
\begin{align*}
\left\| {\rm Res} \left( \widetilde{\Xi}_0 (z) \varphi ; z=e^{i \theta^{(2)}} \right) \right\|^2
&= \frac{(1 - \sqrt{2}S)^2}{2(3 - 2 \sqrt{2}S)^2} \times |\alpha - i \beta|^2, 
\\
\left\| {\rm Res} \left( \widetilde{\Xi}_0 (z) \varphi ; z=e^{i \theta^{(3)}} \right) \right\|^2
&= \frac{(1 - \sqrt{2}S)^2}{2(3 - 2 \sqrt{2}S)^2} \times |\alpha + i \beta|^2,
\\
\left\| {\rm Res} \left( \widetilde{\Xi}_0 (z) \varphi ; z=e^{i \theta^{(4)}} \right) \right\|^2
&=  \frac{(1 - \sqrt{2}S)^2}{2(3 - 2 \sqrt{2}S)^2} \times |\alpha + i \beta|^2.
\end{align*}
Here we used
\begin{align*}
\widetilde{f}_0(e^{i \theta^{(2)}}) = S+Ci, \qquad \widetilde{f}_0(e^{i \theta^{(k)}}) = S-Ci \quad (k=3,4).
\end{align*}
Konno et al. \cite{KLS2013} presented the following key result to obtain the time-averaged limit measure: 
\begin{lem}
\label{yasumiowari}
\begin{align*}
\overline{\mu}_{\infty}(x) = \sum_{k=1}^4 \left\| {\rm Res} \left( \widetilde{\Xi}_x (z) \varphi ; z=e^{i \theta^{(k)}} \right) \right\|^2. 
\end{align*}
\end{lem}
By using Lemma \ref{yasumiowari}, we obtain the time-averaged limit measure at $x=0$:
\begin{lem}
\label{1DkanekuimusiGen}
\begin{align*}
\overline{\mu}_{\infty}(0) = \frac{2(1 - \sqrt{2}S)^2}{(3 - 2 \sqrt{2}S)^2} \times I_{[0,\pi/4)}(\xi).
\end{align*}
\end{lem}

Next, we consider $x \not= 0$ case. To do so, we will compute $\lp (z) =\lp_x(z) \> (x \ge 1)$ for $z=e^{i \theta^{(k)}} \> (k=1,2,3,4)$. We begin with 
\begin{align*}
\lp (e^{i \theta}) 
= \frac{d e^{i \theta} }{1-c \widetilde{f}_0 (e^{i \theta})}
= \frac{- e^{i \theta} }{\sqrt{2} -  \widetilde{f}_0 (e^{i \theta})}.
\end{align*}
From Eq. \eqref{1Dhardnut1Gen}, we have 
\begin{align*}
\lp (e^{i \theta}) 
&= \frac{-1}{\sqrt{2}e^{-i \theta} - ( \sqrt{2} \cos \theta + i {\rm sgn} (\sin \theta) \sqrt{1 - 2 \cos^2 \theta})}
\\
&= - \frac{i}{\sqrt{2} \sin \theta + {\rm sgn} (\sin \theta) \sqrt{1 - 2 \cos^2 \theta}}
\\
&= -i \left(\sqrt{2} \sin \theta - {\rm sgn}(\sin \theta) \sqrt{1 - 2 \cos^2 \theta} \right).
\end{align*}
First, we consider $\theta^{(1)}$ case. Then we get
\begin{align}
\widetilde{\lambda}^{(+)} (e^{i \theta^{(1)}}) = - \frac{i}{\sqrt{3 - 2 \sqrt{2}S}}.
\label{1Dorient1Gen}
\end{align}
Similarly, we see that 
\begin{align}
\widetilde{\lambda}^{(+)} (e^{i \theta^{(1)}}) = - \widetilde{\lambda}^{(+)} (e^{i \theta^{(2)}}) = - \widetilde{\lambda}^{(+)} (e^{i \theta^{(3)}})= \widetilde{\lambda}^{(+)} (e^{i \theta^{(4)}}).
\end{align}
Moreover, noting $\lm (z) = \lm_x (z) \> (x \le -1)$, we have
\begin{align*}
\lm (e^{i \theta}) 
= \frac{a e^{i \theta} }{1-b \widetilde{f}_0 (e^{i \theta})}
= \frac{e^{i \theta} }{\sqrt{2} -  \widetilde{f}_0 (e^{i \theta})}
= - \lp (e^{i \theta}).
\end{align*}
So we get
\begin{align}
\widetilde{\lambda}^{(-)} (e^{i \theta^{(k)}}) = - \widetilde{\lambda}^{(+)} (e^{i \theta^{(k)}}) \quad (k=1,2,3,4).
\end{align}
Therefore, we obtain
\begin{lem}
\label{1DkanekuiGen}
\begin{align*}
\left| \widetilde{\lambda}^{(\pm)} (e^{i \theta^{(k)}}) \right|^2 = \frac{1}{3-2 \sqrt{2}S} \quad (k=1,2,3,4).
\end{align*}
\end{lem}

From now on, we consider $x \not=0$ case. First, we compute $x \ge 1$ case. From Proposition \ref{acco}, we have
\begin{align} 
\widetilde{\Xi}_x(z) \varphi
&= \lp_{x-1}(z) \cdots \lp_{1}(z)
\begin{bmatrix} 
\lp_{x}(z)\fp_x(z) \\ z
\end{bmatrix}
\begin{bmatrix} c_0 & d_0 
\end{bmatrix}
\widetilde{\Xi}_0(z) \varphi
\nonumber
\\
&= (\lp (z))^{x-1} 
\begin{bmatrix} 
\lp (z) \widetilde{f}_0(z) \\ z
\end{bmatrix}
\begin{bmatrix} 
S & -C
\end{bmatrix}
\widetilde{\Xi}_0(z) \varphi
\nonumber
\\
&=
(\lp (z))^{x-1} 
\begin{bmatrix} 
S \lp (z) \widetilde{f}_0(z)
& - C \lp (z) \widetilde{f}_0(z)
\\
Sz 
& - Cz
\end{bmatrix}
\nonumber
\\
& \qquad \qquad \qquad \times 
\frac{1}{\gamma (z)}
\begin{bmatrix} 
(1 - S \widetilde{f}_0(z)) \alpha - C \widetilde{f}_0(z) \beta
\\ 
C \widetilde{f}_0(z) \alpha + (1 - S \widetilde{f}_0(z)) \beta
\end{bmatrix}.
\label{1Dpakepake1Gen}
\end{align}
In order to compute the residue, we introduce 
\begin{align*}
I_1 (z)
&= 
\dfrac{(\lp (z))^{x}\widetilde{f}_0(z)}{\gamma' (z)} 
\left\{ (S - \widetilde{f}_0(z)) \alpha - C \beta \right\},
\\
I_2 (z)
&= 
\dfrac{z (\lp (z))^{x-1}}{\gamma' (z)} 
\left\{ (S - \widetilde{f}_0(z)) \alpha - C \beta \right\}.
\end{align*}
First, we consider $z=e^{i \theta^{(1)}}$. Then, we get
\begin{align}
|I_1 (e^{i \theta^{(1)}})|^2
&= 
\dfrac{|\lp (e^{i \theta^{(1)}})|^{2x}}{|\gamma' (e^{i \theta^{(1)}})|^2}
\left|\widetilde{f}_0(e^{i \theta^{(1)}})\right|^2  
\left|  (S - \widetilde{f}_0(e^{i \theta^{(1)}})) \alpha - C \beta 
\right|^2
\nonumber
\\
&= 
\frac{(1- \sqrt{2}S)^2}{4(3-2\sqrt{2}S)^2} \left( \frac{1}{3-2\sqrt{2}S} \right)^x \left|\alpha - i \beta \right|^2.
\label{1Dpakepake2Gen}
\end{align}
Here we used $\widetilde{f}_0(e^{i \theta^{(1)}})=S+Ci$. Similarly, we have
\begin{align}
|I_2 (e^{i \theta^{(1)}})|^2
= 
\frac{(1- \sqrt{2}S)^2}{4(3-2\sqrt{2}S)^2} \left( \frac{1}{3-2\sqrt{2}S} \right)^{x-1} \left|\alpha - i \beta \right|^2.
\label{1Dpakepake3Gen}
\end{align}
By using Eqs. \eqref{1Dpakepake1Gen}, \eqref{1Dpakepake2Gen}, and \eqref{1Dpakepake3Gen}, we see that for $x \ge 1$, 
\begin{align*} 
\left\| {\rm Res} \left( \widetilde{\Xi}_x (z) \varphi ; z=e^{i \theta^{(1)}} \right) \right\|^2
&= |I_1 (e^{i \theta^{(1)}})|^2 + |I_2 (e^{i \theta^{(1)}})|^2
\\
&= \frac{(2- \sqrt{2}S)(1- \sqrt{2}S)^2}{2(3-2\sqrt{2}S)^2} \left( \frac{1}{3-2\sqrt{2}S} \right)^x \left|\alpha - i \beta \right|^2.
\end{align*}
Next, we consider $z=e^{i \theta^{(2)}}$ case. In a similar fashion, we get
\begin{align*} 
\left\| {\rm Res} \left( \widetilde{\Xi}_x (z) \varphi ; z=e^{i \theta^{(2)}} \right) \right\|^2
=\frac{(2- \sqrt{2}S)(1- \sqrt{2}S)^2}{2(3-2\sqrt{2}S)^2} \left( \frac{1}{3-2\sqrt{2}S} \right)^x \left|\alpha - i \beta \right|^2.
\end{align*}
For $z=e^{i \theta^{(k)}} \> (k=3,4)$ cases also, we have
\begin{align*} 
\left\| {\rm Res} \left( \widetilde{\Xi}_x (z) \varphi ; z=e^{i \theta^{(k)}} \right) \right\|^2
=
\frac{(2- \sqrt{2}S)(1- \sqrt{2}S)^2}{2(3-2\sqrt{2}S)^2} \left( \frac{1}{3-2\sqrt{2}S} \right)^x \left|\alpha + i \beta \right|^2.
\end{align*}
Therefore, we obtain
\begin{align*}
\sum_{k=1}^4 \left\| {\rm Res} \left( \widetilde{\Xi}_x (z) \varphi ; z=e^{i \theta^{(k)}} \right) \right\|^2
= \frac{2(2- \sqrt{2}S)(1- \sqrt{2}S)^2}{(3-2\sqrt{2}S)^2} \left( \frac{1}{3-2\sqrt{2}S} \right)^x.
\end{align*}
Thus, for $x \ge 1$, we see that Lemma \ref{yasumiowari} implies
\begin{align*}
\overline{\mu}_{\infty}(x) = \frac{2(2- \sqrt{2}S)(1- \sqrt{2}S)^2}{(3-2\sqrt{2}S)^2} \left( \frac{1}{3-2\sqrt{2}S} \right)^x \times I_{[0,\pi/4)}(\xi).
\end{align*}

Next, we consider $x \le -1$ case. By Proposition \ref{acco}, we have
\begin{align} 
\widetilde{\Xi}_x(z) \varphi
&= \lm_{x+1}(z) \cdots \lm_{-1}(z)
\begin{bmatrix} 
z \\ \lm_{x}(z)\fm_x(z) 
\end{bmatrix}
\begin{bmatrix} a_0 & b_0 
\end{bmatrix}
\widetilde{\Xi}_0(z) \varphi
\nonumber
\\
&= (\lm (z))^{|x|-1} 
\begin{bmatrix} 
z \\ \lm (z) \widetilde{f}_0(z) 
\end{bmatrix}
\begin{bmatrix} C & S
\end{bmatrix}
\widetilde{\Xi}_0(z) \varphi
\nonumber
\\
&=
(\lm (z))^{|x|-1} 
\begin{bmatrix} 
 C z
& S z 
\\
C \lm (z) \widetilde{f}_0(z)
& S \lm (z) \widetilde{f}_0(z)
\end{bmatrix}
\nonumber
\\
& \qquad \qquad \qquad \times 
\frac{1}{\gamma (z)}
\begin{bmatrix} 
(1 - S \widetilde{f}_0(z)) \alpha - C \widetilde{f}_0(z) \beta
\\ 
C \widetilde{f}_0(z) \alpha + (1 - S \widetilde{f}_0(z)) \beta
\end{bmatrix}.
\label{1Dpakopako1Gen}
\end{align}
In order to calculate the residue, we introduce 
\begin{align*}
J_1 (z)
&= 
\dfrac{z (\lm (z))^{|x|-1}}{\gamma' (z)} 
\left\{ C \alpha + (S - \widetilde{f}_0(z)) \beta \right\},
\\
J_2 (z)
&= 
\dfrac{(\lm (z))^{|x|}\widetilde{f}_0(z)}{\gamma' (z)} 
\left\{ C \alpha + (S - \widetilde{f}_0(z)) \beta \right\}.
\end{align*}
First, we consider $z=e^{i \theta^{(1)}}$ case. Then, we get 
\begin{align}
|J_1 (e^{i \theta^{(1)}})|^2
= 
\frac{(1- \sqrt{2}S)^2}{4(3-2\sqrt{2}S)^2} \left( \frac{1}{3-2\sqrt{2}S} \right)^{|x|-1} \left|\alpha - i \beta \right|^2.
\label{1Dpakopako2Gen}
\end{align}
Similarly, we have
\begin{align}
|J_2 (e^{i \theta^{(1)}})|^2
= 
\frac{(1- \sqrt{2}S)^2}{4(3-2\sqrt{2}S)^2} \left( \frac{1}{3-2\sqrt{2}S} \right)^{|x|} \left|\alpha - i \beta \right|^2.
\label{1Dpakopako3Gen}
\end{align}
Thus, by using Eqs. \eqref{1Dpakopako1Gen}, \eqref{1Dpakopako2Gen}, and \eqref{1Dpakopako3Gen}, we see that for $x \le -1$, 
\begin{align*} 
\left\| {\rm Res} \left( \widetilde{\Xi}_x (z) \varphi ; z=e^{i \theta^{(1)}} \right) \right\|^2
&= |J_1 (e^{i \theta^{(1)}})|^2 + |J_2 (e^{i \theta^{(1)}})|^2
\\
&= \frac{(2- \sqrt{2}S)(1- \sqrt{2}S)^2}{2(3-2\sqrt{2}S)^2} \left( \frac{1}{3-2\sqrt{2}S} \right)^{|x|} \left|\alpha - i \beta \right|^2.
\end{align*}
For $z=e^{i \theta^{(2)}}$ case, we similarly have
\begin{align*} 
\left\| {\rm Res} \left( \widetilde{\Xi}_x (z) \varphi ; z=e^{i \theta^{(2)}} \right) \right\|^2
= 
\frac{(2- \sqrt{2}S)(1- \sqrt{2}S)^2}{2(3-2\sqrt{2}S)^2} \left( \frac{1}{3-2\sqrt{2}S} \right)^{|x|} \left|\alpha - i \beta \right|^2.
\end{align*}
For $z=e^{i \theta^{(k)}} \> (k=3,4)$ cases also, we get
\begin{align*} 
\left\| {\rm Res} \left( \widetilde{\Xi}_x (z) \varphi ; z=e^{i \theta^{(k)}} \right) \right\|^2
= \frac{(2- \sqrt{2}S)(1- \sqrt{2}S)^2}{2(3-2\sqrt{2}S)^2} \left( \frac{1}{3-2\sqrt{2}S} \right)^{|x|} \left|\alpha + i \beta \right|^2.
\end{align*}
So, we obtain
\begin{align*}
&\sum_{k=1}^4 \left\| {\rm Res} \left( \widetilde{\Xi}_x (z) \varphi ; z=e^{i \theta^{(k)}} \right) \right\|^2
\\
& \qquad = \frac{2(2- \sqrt{2}S)(1- \sqrt{2}S)^2}{(3-2\sqrt{2}S)^2} \left( \frac{1}{3-2\sqrt{2}S} \right)^{|x|}.
\end{align*}
Therefore, we see that for $x \le -1$, Lemma \ref{yasumiowari} gives
\begin{align*}
\overline{\mu}_{\infty}(x) = \frac{2(2- \sqrt{2}S)(1- \sqrt{2}S)^2}{(3-2\sqrt{2}S)^2} \left( \frac{1}{3-2\sqrt{2}S} \right)^{|x|}.
\end{align*}
Then we have $\overline{\mu}_{\infty}(x) = \overline{\mu}_{\infty}(-x) \> (x \in \ZM)$.

Combaining Lemma \ref{1DkanekuimusiGen} with the above result gives
\begin{thm}
\label{sanosan2Gen} 
\begin{align*}
\overline{\mu}_{\infty} (x) = 
\left\{ \begin{array}{ll}
\dfrac{2(1- \sqrt{2}S)^2}{(3-2\sqrt{2}S)^2} \times I_{[0,\pi/4)} (\xi) & (x=0), \\
\\
\dfrac{2(2- \sqrt{2}S)(1- \sqrt{2}S)^2}{(3-2\sqrt{2}S)^2} \left( \dfrac{1}{3-2\sqrt{2}S} \right)^{|x|} \times I_{[0,\pi/4)} (\xi) & (x \neq 0), 
\end{array} 
\right.
\end{align*}
and
\begin{align*}
\sum_{x \in \ZM} \overline{\mu}_{\infty} (x) = \frac{2(1- \sqrt{2}S)}{3- 2 \sqrt{2}S} \times I_{[0,\pi/4)} (\xi) (<1).
\end{align*}
\end{thm}
The time-avereged limit measure does not depend on the inital qubit $\varphi = {}^T [\alpha, \beta] \> (\alpha, \beta \in \CM, |\alpha|^2 + |\beta|^2 =1)$. Furthermore, we take $|c|=\sqrt{2(1- \sqrt{2}S)/(3- 2 \sqrt{2}S)}$ in Corollary \ref{yasumihayai} on the stationary measure and have the same result.

\par
\
\par\noindent
{\bf Acknowledgments.} NK acknowledges financial supports of the Grant-in-Aid for Scientific Research (C) from Japan Society for the Promotion of Science (Grant No.24540116). ES thanks to the financial support of the Grant-in-Aid for Young Scientists (B) of Japan Society for the Promotion of Science (Grant No.25800088).

\par
\
\par

\begin{small}
\bibliographystyle{jplain}

\end{small}

\end{document}